\documentclass[12pt]{article}

\usepackage[T1]{fontenc}
\usepackage{mathptmx}
\usepackage{microtype}
\usepackage[a4paper,margin=1in]{geometry}
\usepackage{setspace}
\usepackage{amsmath,amssymb,amsthm,mathtools}

\usepackage{graphicx,float}
\usepackage{booktabs,array,multirow,longtable}
\usepackage[round,authoryear]{natbib}
\usepackage[dvipsnames]{xcolor}
\definecolor{brandburgundy}{RGB}{128,0,32}
\usepackage{url,hyperref}
\usepackage{enumitem,fancyhdr,tcolorbox}
\usepackage{listings}
\usepackage{tikz}
\usetikzlibrary{arrows.meta,positioning,shapes.geometric}
\usepackage{pgfplots}
\pgfplotsset{compat=1.18}

\lstdefinelanguage{Lean}{
  morekeywords={theorem,lemma,def,noncomputable,import,open,namespace,end,
    by,simp,rw,ring,linarith,exact,have,intro,obtain,refine,apply,
    fun,let,where,if,then,else,match,with,section,variable,class,
    instance,structure,inductive,axiom,sorry,calc,show,suffices,
    unfold,rcases,positivity,nlinarith},
  sensitive=true,
  morecomment=[l]{--},
  morecomment=[n]{/-}{-/},
  morestring=[b]",
  literate=
    {ℝ}{{$\mathbb{R}$}}1 {ℂ}{{$\mathbb{C}$}}1 {ℕ}{{$\mathbb{N}$}}1
    {→}{{$\to$}}1 {←}{{$\leftarrow$}}1 {↔}{{$\leftrightarrow$}}1
    {≤}{{$\le$}}1 {≥}{{$\ge$}}1 {≠}{{$\ne$}}1 {¬}{{$\lnot$}}1
    {∀}{{$\forall$}}1 {∃}{{$\exists$}}1 {∈}{{$\in$}}1 {∉}{{$\notin$}}1
    {⟨}{{$\langle$}}1 {⟩}{{$\rangle$}}1 {▸}{{$\blacktriangleright$}}1
    {·}{{$\cdot$}}1 {×}{{$\times$}}1 {⊢}{{$\vdash$}}1
    {λ}{{$\lambda$}}1 {↦}{{$\mapsto$}}1 {⟹}{{$\Longrightarrow$}}1
}
\theoremstyle{definition}
\newtheorem{definition}{Definition}[section]

\theoremstyle{plain}
\newtheorem{theorem}{Theorem}[section]

\newtheorem{observation}[theorem]{Observation}
\theoremstyle{remark}

\newcommand{\E}{\mathbb{E}}
\newcommand{\TSpointer}[1]{%
  \textnormal{\textsc{#1}}}

\def\UrlBreaks{\do\/\do-\do_}
\expandafter\def\expandafter\UrlBreaks\expandafter{\UrlBreaks%
  \do\a\do\b\do\c\do\d\do\e\do\f\do\g\do\h\do\i\do\j\do\k%
  \do\l\do\m\do\n\do\o\do\p\do\q\do\r\do\s\do\t\do\u\do\v%
  \do\w\do\x\do\y\do\z}

\usepackage{titlesec}
\titleformat{\section}{\normalfont\large\bfseries\color{brandburgundy}}{\thesection}{0.5em}{}
\titleformat{\subsection}{\normalfont\normalsize\bfseries\color{brandburgundy}}{\thesubsection}{0.5em}{}
\titleformat{\subsubsection}{\normalfont\small\bfseries\color{brandburgundy}}{\thesubsubsection}{0.5em}{}
\titlespacing*{\section}{0pt}{2ex plus 0.8ex minus 0.2ex}{1ex plus 0.3ex}
\titlespacing*{\subsection}{0pt}{1.5ex plus 0.5ex minus 0.2ex}{0.8ex plus 0.2ex}
\titlespacing*{\subsubsection}{0pt}{1.2ex plus 0.4ex minus 0.2ex}{0.6ex plus 0.2ex}

\fancypagestyle{firstpage}{\fancyhf{}\fancyfoot[C]{\small\thepage}%
  }

\newcommand{\papernum}{DAI-2601}
\newcommand{\paperstatus}{Preprint}
\newcommand{\paperdate}{24 August 2026}
\newcommand{\paperdoi}{10.48550/arXiv.2601.06692}
\newcommand{\paperurl}{https://systems.ac/1/DAI-2601}

\hypersetup{colorlinks=true,linkcolor=brandburgundy,citecolor=brandburgundy,
  urlcolor=brandburgundy,breaklinks=true,
  pdftitle={The Axiom of Consent: Authorization, Friction, and Multi-Agent Coordination},
  pdfauthor={Murad Farzulla},
  pdfkeywords={authorization, multi-agent systems, coordination, friction, legitimacy, measurement}}

\begin{document}
\setstretch{1.6}
\thispagestyle{firstpage}

\begin{center}
{\small\textsc{\href{https://dissensus.ai}{Dissensus} Working Paper Series}}\\[0.2em]
{\small \href{\paperurl}{\papernum}}
\end{center}
\vspace{1.2em}

\begin{center}
{\LARGE\bfseries The Axiom of Consent:}\\[0.3em]
{\large\itshape Authorization, Friction, and Multi-Agent Coordination}\\[1.3em]
{\large Murad Farzulla}\textsuperscript{1,2,*}\\[0.7em]
{\small
  \textsuperscript{1}\href{https://dissensus.ai}{Dissensus}, London, UK \quad
  \textsuperscript{2}King's College London, London, UK}\\[0.4em]
{\footnotesize
  \textsuperscript{*}Correspondence: \href{mailto:murad@dissensus.ai}{murad@dissensus.ai}
  \quad ORCID: \href{https://orcid.org/0009-0002-7164-8704}{0009-0002-7164-8704}}\\[0.3em]
{\footnotesize \paperdate}
\end{center}

\vspace{0.4em}
\begin{abstract}
Coordination research collapses four objects: operative control, authorization, a model-derived friction score, and observed outcomes. The Axiom of Consent is a stake-weighted unanimity principle; majority and supermajority thresholds are explicit relaxations, not versions of the axiom. Decision loci are structural facts, whereas authorization and legitimacy require normative and measurement premises.

Alignment, calibrated stakes, and information deficit are candidate coordinates, and $F=\sigma(1+\varepsilon)/(1+\alpha)$ is a phenomenological ansatz. The Replicator--Optimization Mechanism supplies a conditional persistence interface: irreducibility suffices for its finite, static, positive-fitness continuous-time Perron result; primitivity is required only for the corresponding discrete-time power convergence, and the componentwise ranking is narrower. Neither persistence result derives authorization.

A resource-allocation instantiation specifies an identification contract but observes no authorization acts or effective voice. Its exploratory MARL companion uses target-vector correlation and observation noise as narrow proxy treatments, not measures of general alignment or information deficit. Under that proxy and reward-gap design, the composite loses to an independent-effects model and a feasible-centred frozen crossing yields the opposite interaction direction. Cooperative target correlation lowers the gap under shared-state contention; separable IQL is structurally invariant and separable VDN is a non-detection. Paired partial sharing modulates the gradient without establishing an exact dose law or endpoint equivalence. Target support changes opposition and residual-policy conclusions. The surviving contribution is an authorization architecture and measurement discipline, not a universal friction law.

\end{abstract}

\vspace{0.6em}
\noindent{\footnotesize
  \textbf{DOI:} \href{https://doi.org/\paperdoi}{\texttt{\paperdoi}}
  \quad|\quad \textbf{Status:} \paperstatus
  \quad|\quad \textbf{License:} CC-BY-4.0}

\smallskip
\noindent{\footnotesize\textbf{Ancillary supplement:}
\texttt{anc/technical-supplement\_AOC.pdf}. Pointers labelled \textsc{TS-*} resolve to named
destinations within that separately downloadable PDF.}

\section{The layer-separation problem}
\label{sec:layers}

Multi-agent coordination joins questions that are easy to state and dangerous to merge. Who can
select an outcome? Who ought to be able to select it? How should disagreement, exposure, and
information be represented? Which observable outcomes would count against the resulting model?
A protocol, market, committee, or learned policy can answer the first question while remaining
silent on the other three. Conversely, a normative principle can specify conditions of
authorization without predicting deadlock, reward loss, exit, or persistence. The central claim
of this paper is architectural: those objects belong to different layers and should be linked by
explicit bridges rather than by vocabulary.

The distinction matters because low resistance can arise from many causes. A decision may be
widely authorized; affected parties may lack an exit; opposition may be costly to express; the
system may attenuate observable consequences; or the measured outcome may simply be insensitive
to the conflict under study. Calling every low-outcome configuration ``consensual'' turns an
empirical and normative question into a definition. The reverse error is also common: observable
conflict is treated as proof that authorization was absent, although an authorized decision can
remain costly to implement. The framework therefore begins with four layers.

\begin{table}[htbp]
\centering
\caption{Objects that the framework keeps separate. Each proposed link between rows is a claim
requiring its own argument or evidence.}
\label{tab:layers}
\begin{tabular}{@{}p{2.7cm}p{4.6cm}p{6.0cm}@{}}
\toprule
Layer & Object & Question and failure condition \\
\midrule
Operative & Decision locus or controller & Which procedure actually selects the outcome? A
misspecified locus invalidates all downstream measures. \\
Authorization & Stake-weighted consent predicate & Was that locus authorized by affected agents?
Control alone supplies no affirmative answer. \\
Model & Candidate coordinates and friction index & Does a declared function summarize the
measured inputs? Functional comparison can reject the index without settling authorization. \\
Outcome & Separately specified, index-distinct $Y$ & Does the index predict coordination loss,
resistance, or persistence out of sample? Low $Y$ is not consent by definition. \\
\bottomrule
\end{tabular}
\end{table}

This layered view changes the role of the proposed expression
\begin{equation}
F=\sigma\frac{1+\varepsilon}{1+\alpha}.
\label{eq:canonical-intro}
\end{equation}
Alignment $\alpha$, calibrated stake $\sigma$, and information deficit $\varepsilon$ are candidate
coordinates for a specified decision domain. Equation~\eqref{eq:canonical-intro} is a
phenomenological ansatz: it supplies transparent comparative statics and a falsifiable composite,
not a consequence of the authorization principle. In its exploratory target-correlation and
observation-noise proxy designs, the companion control battery does not support it as a single
predictor of the specified reward-gap outcome. That result is part of the framework's empirical
status, not an inconvenience to be hidden. It leaves open whether individual coordinates, a
different composition, or a different outcome model can be useful under narrower scope
conditions.

\subsection{Decision loci before consent-holders}

Whenever a non-null outcome is selected in a shared domain, some procedure determines it. The
procedure may be concentrated in an executive, distributed across a vote, encoded in a contract,
or delegated to randomization. This is a structural observation about selection. It says neither
that the procedure has a right to decide nor that affected parties consented to its use. We call
the procedure an \emph{operative decision locus}. The term \emph{consent-holder} is reserved for
an operative locus that also satisfies the authorization conditions stated in
Section~\ref{sec:authorization}.

The distinction avoids two equivocations. First, a market, algorithm, or inherited convention is
not ownerless simply because no person selects each outcome by hand. A prior rule allocates
control, and that allocation can be studied. Second, locating control does not confer legitimacy.
A system can be operationally decisive and normatively unauthorized. Resource permissions,
agenda control, vetoes, default rules, and model-deployment rights are evidence about operative
control; none is, by itself, evidence of consent.

Mechanism design offers a natural comparison. It studies allocation rules, incentive compatibility,
and preference revelation with remarkable precision
\citep{hurwicz1960optimality,myerson1981optimal,maskin1999nash}. The present question begins one
step earlier and ends one step later: which affected agents enter the authorization set, how are
their stakes made commensurable, and does a model-derived index predict a separately specified
outcome after shared algebra has been audited? These questions complement rather than replace mechanism design.
Communication and delegation research likewise show how commitment and hierarchy can change
coordination efficiency \citep{avoyan2023road,kocak2023dual}; efficiency does not settle the
authorization of the arrangement producing it.

\subsection{Normative authorization and descriptive modeling}

The informal Axiom of Consent is demanding: no affected entity should be bound by a commitment it
did not authorize, with standing weighted by its stake in the outcome. Read literally, this is a
unanimity principle. It does not follow from the fact that systems need decision rules, and it
does not become true because a low-friction arrangement persists. Section~\ref{sec:authorization}
formalizes the principle as the $\theta=1$ case of a threshold rule. Values $\theta<1$ allow the
model to represent majority or supermajority procedures, but they are explicit relaxations. The
threshold parameter cannot quietly transform a unanimity axiom into a family of equally
authorized rules.

The descriptive apparatus makes different commitments. It proposes measurements for controller--
stakeholder alignment, exposure, and information loss; a function combining them; a separately
specified, index-distinct outcome; and, conditionally, a persistence process. Each commitment can
fail separately. Objective
correlation may not measure consent. Raw utility gradients may lack a common unit. Information
loss may interact with alignment in the opposite direction from the ratio. A fitted index may not
predict the chosen outcome. A survival law may fail even if the cross-sectional index performs
well. The framework is useful only if those failures remain visible.

This separation also blocks a naturalistic shortcut. Replicator dynamics describe differential
persistence under a specified fitness map \citep{taylor1978evolutionary,price1970selection}; they
do not show that what survives is legitimate. Evolutionary accounts of conventions and social
contracts demonstrate how norm-like arrangements can be selected
\citep{skyrms1996evolution,young1993evolution,binmore1994playing,binmore1998just}. The conditional
ROM interface in Section~\ref{sec:rom-interface} follows that explanatory strategy, but its
legitimacy term is a stipulated bridge. Survival remains evidence about persistence under the
model, never a derivation of moral authority.

\subsection{Why multi-agent resource allocation is the principal test bed}

Multi-agent resource allocation makes all four layers inspectable. Access controls and transition
rules reveal who can move shared resources. Reward or utility functions provide a declared, if
imperfect, representation of objectives. Stakes can be tied to a specified allocation increment.
Communication channels and observation corruption can be manipulated. Deadlock, coordination
gap, reward loss, negotiation cost, and policy performance can be specified separately from the
index, while any shared reward algebra is audited explicitly. MARL also makes the failure modes sharp: decentralized control is hard in general
\citep{bernstein2002complexity}, reward functions are often taken as given
\citep{busoniu2008comprehensive,zhang2021multi}, and a treatment named ``opposition'' may not
actually deliver negative objective correlation inside the feasible state space.

The companion study therefore does more than supply a favorable example. Its exploratory controls expose four ways a
coordination finding can be manufactured: a sign-blind preference sampler, a scale factor built
into reward algebra, a shared-state mechanism mistaken for a universal alignment law, and a
training-window gap mistaken for learned-policy quality. In that target-coordinate proxy
instantiation, repairs to those artifacts reject the composite as a reward-gap predictor while
preserving a narrower, scope-conditioned association. Section
\ref{sec:empirical-status} treats the nulls, contradictions, and support sensitivity as findings of
record.

\subsection{Where the framework sits}

The framework intersects four literatures but claims a different object from each. The first is
the literature on collective authorization. Social-choice theory asks how individual orderings can
be aggregated and proves that attractive properties cannot generally be satisfied together
\citep{arrow1951social}. Democratic boundary theory asks who is entitled to participate when a
decision affects people beyond a formal jurisdiction
\citep{goodin2007enfranchising,arrhenius2005boundary}. Contractualist theories ask what principles
could survive reasonable rejection \citep{scanlon1998we}. AOC does not solve the aggregation
impossibilities or supply a complete theory of reasonableness. Its narrower move is to put the
affected set, stake weights, verified authorization act, and operative decision locus into separate
positions in an empirical specification. This makes it possible to see whether a study has measured
authorization or merely favorable outcomes.

The second is institutional and mechanism design. Mechanisms transform messages into outcomes and
can make truth-telling or efficient allocation incentive compatible
\citep{hurwicz1960optimality,myerson1981optimal,maskin1999nash}. Polycentric governance research
shows that durable common-pool arrangements often rely on locally fitted rules, monitoring, and
graduated sanctions rather than one universal architecture \citep{ostrom1990governing}. These
traditions provide tools for designing the operative locus. AOC adds two audits. It asks whether
the rule was authorized by the agents assigned positive stakes and whether the outcome used to
evaluate it is separately specified from the score used to design it and any shared algebra has
been audited. An incentive-compatible rule can still
exclude affected outsiders; an authorized rule can still perform badly; and a locally successful
institution does not validate a cross-domain scalar.

The third is the evolutionary social-contract tradition. Replicator dynamics and stochastic
adaptive play explain how strategies or conventions gain population share
\citep{taylor1978evolutionary,young1993evolution}. Skyrms and Binmore show that fairness-like
conventions can arise through selection rather than hypothetical agreement
\citep{skyrms1996evolution,binmore1994playing,binmore1998just}. The present interface selects over
control configurations, and it makes effective voice, interpreted as legitimacy only through a
declared bridge, and a model index visible in an effective-fitness map. That visibility is useful
precisely because it reveals the bridge premise. If one posits that
effective voice improves persistence, the model can trace the implications. Selection does not
establish the premise, and prevalence does not authorize a configuration.

The fourth is multi-agent learning and cooperative AI. MARL studies policies that interact through
a shared environment, often under common, individual, or mixed rewards
\citep{leibo2017multi,lerer2019learning,zhang2021multi}. Value decomposition supplies a way to train
decentralized policies from a team objective \citep{sunehag2018value}; reward-mixing approaches can
change the strategic game itself \citep{gemp2022d3c}. These methods show that objective structure
matters, but they do not make reward correlation identical to consent. They also highlight a
measurement problem: an apparent coordination benefit can be produced by reward scale, target
feasibility, state coupling, or evaluation noise. AOC's contribution is not a new learning
algorithm. It is the contract requiring those mechanisms and estimands to be separated before a
result is assigned normative meaning.

This positioning explains why the paper does not seek a maximally comprehensive synthesis. A
single framework cannot inherit the strongest conclusions of each neighboring theory merely by
sharing words such as preference, voice, fitness, or coordination. The relevant interfaces are
typed. Mechanism design informs the operative rule; authorization theory defines the normative
test; measurement theory defines the constructs; MARL supplies a controllable empirical domain;
and ROM supplies a conditional persistence model. Evidence crosses an interface only through a
declared mapping.

\subsection{From a broad ambition to a testable programme}

The long-form edition treated recurring resistance across political, financial, computational,
and organizational systems as motivation for a common description. That remains a possible
research programme, but common language is not evidence of a common latent variable. Exit in a
market, policy reversal in an institution, reward loss in an MDP, and protest in a polity differ in
agency, time scale, observability, and causal structure. A cross-domain claim would require at
least configural, metric, and scalar invariance: the same construct must be present, changes in its
measure must have comparable meaning, and levels must be comparable after calibration. None is
established here.

The reader paper therefore uses multi-agent resource allocation as the single worked domain. This
choice is not based on the domain being morally simple. It is based on experimental visibility.
The transition map, reward support, action constraints, communication channel, and evaluation
policy can be inspected. Counterfactual controls can hold latent problems fixed while changing
shared coordinates. An exact dynamic oracle can be computed in the chosen environment. A failure
under these favorable measurement conditions is especially informative: it warns against carrying
the composite into domains where the constructs are less observable.

The broader programme can proceed incrementally. A new domain should first validate its own
affected set, locus, stake scale, alignment measure, information channel, and outcome. It should
then test whether the coordinate measures are invariant across relevant groups or time periods.
Only after within-domain prediction and calibration succeed should a shared functional form be
compared across domains. This sequence treats transfer as a hypothesis, not as a benefit conferred
by notation.

\subsection{Contribution and boundary}

The reader-facing contribution is narrower than the earlier long-form edition. First, it specifies
a stake-weighted authorization rule while keeping operative control, authorization, legitimacy,
model index, and outcomes distinct. Second, it supplies a measurement contract for any empirical
instantiation: define the locus and affected set, normalize exposure, declare the objective and
state sampling distribution, define the information channel, and prespecify an index-distinct
outcome with a common-method audit. Third, it gives a formal multi-agent application and reports the current, adverse verdict
on the proposed composite. Fourth, it exposes a short, conditional interface to the
Replicator--Optimization Mechanism (ROM) rather than reproducing ROM's persistence theory.

Detailed rational-inattention motivations, information-decomposition analogies, proxy catalogues,
transfer-entropy procedures, alternative causal designs, and cross-domain illustrations remain in
the checksum-identified long technical record. ROM-owned scale, ownership, and persistence results
live in the ROM reader and supplement. They are navigable through pointers \TSpointer{TS-FORM},
\TSpointer{TS-ROM}, \TSpointer{TS-MEASURE}, and \TSpointer{TS-DOMAIN}. This division does not demote
their possible use; it prevents them from obscuring the paper's load-bearing claims. The paper proceeds from the
authorization rule, through a conditional persistence interface, to the multi-agent
instantiation, measurement contract, empirical verdict, and the objections that delimit what the
framework can say.

\section{Authorization and the core concepts}
\label{sec:authorization}

\subsection{The axiom and its formal relaxation}

\begin{tcolorbox}[colback=brandburgundy!5,colframe=brandburgundy,
title={\textbf{The Axiom of Consent}}]
\emph{No entity should be bound by a commitment it did not authorize, with its standing weighted
by its stake in the outcome.}
\end{tcolorbox}

The axiom is a normative principle, not a statistical regularity. It specifies a demanding boundary
for legitimate commitment: all affected parties with positive stakes must authorize the decision.
The framework uses an affected-interests conception of standing
\citep{goodin2007enfranchising,arrhenius2005boundary} and a proportionality intuition: greater
consequence exposure carries greater weight. Neither commitment is normatively neutral. A study
must defend its affected-set rule and its stake scale rather than let a convenient dataset decide
who counts.

Let $\mathcal A$ be the agents, $\mathcal D$ the decisions, $s_i(d)\geq0$ agent $i$'s calibrated
stake in decision $d$, and $S_d=\{i:s_i(d)>0\}$ the affected set. A thresholded rule is useful for
representing institutions that do not use unanimity.

\begin{definition}[Thresholded authorization rule]
\label{def:authorization}
For a decision $d$ and threshold $\theta\in(0,1]$, define
\begin{equation}
\operatorname{Authorized}_{\theta}(d)
\iff
\sum_{i\in S_d}s_i(d)\,\mathbb 1[\operatorname{Consent}_i(d)]
\geq
\theta\sum_{i\in S_d}s_i(d).
\label{eq:authorization}
\end{equation}
The Axiom of Consent is the case $\theta=1$. Every $\theta<1$ is a modeling relaxation for a
majority or supermajority procedure and does not satisfy the literal unanimity principle.
\end{definition}

This formulation makes three choices explicit. \emph{Affected-set restriction} excludes agents
assigned zero stake, so boundary errors can become authorization errors. \emph{Stakes weighting}
lets greater exposure carry greater influence, but only after calibration has made stakes
commensurable. \emph{Threshold structure} distinguishes the axiom from implementable decision
rules. Work on constitutional choice explains why institutions may select different thresholds
for different decisions \citep{buchanan1962calculus}; it does not make those thresholds instances
of unanimity.

The binary indicator is deliberately austere. It can encode a verified authorization act, but it
does not distinguish enthusiastic endorsement, reluctant acceptance, or consent produced by an
unacceptable option set. Those issues belong to the measurement and normative specification. A
continuous voice score is useful descriptively, but it should not be relabeled consent without an
explicit correspondence.

\begin{definition}[Stake-weighted effective voice]
\label{def:legitimacy}
Assume the affected set is nonempty and
$\sigma(d)=\sum_{i\in S_d}s_i(d)>0$. Let $v_i(d,t)\in[0,1]$ measure agent $i$'s effective
influence on the selected outcome. Define
\begin{equation}
L(d,t)=
\frac{\sum_{i\in S_d}s_i(d)v_i(d,t)}{\sum_{i\in S_d}s_i(d)}.
\label{eq:legitimacy}
\end{equation}
$L$ is a descriptive stakes-weighted effective-voice score. Treating it as a legitimacy measure
requires the declared normative and measurement bridge. When
$v_i=\mathbb 1[\operatorname{Consent}_i]$, the event $L\geq\theta$ reproduces
Equation~\eqref{eq:authorization}; with graded voice it is a relaxation, not a derivation.
\end{definition}

\begin{observation}[Unanimity boundary and continuous relaxation]
\label{obs:unanimity-boundary}
For $\sigma(d)=\sum_{i\in S_d}s_i(d)>0$, define the consenting stake share
\begin{equation}
q(d)=\frac{1}{\sigma(d)}
\sum_{i\in S_d}s_i(d)\mathbb 1[\operatorname{Consent}_i(d)].
\label{eq:consenting-share}
\end{equation}
Then $\operatorname{Authorized}_\theta(d)$ is equivalent to $q(d)\geq\theta$. At
$\theta=1$, positive stakes imply $q(d)=1$ if and only if every $i\in S_d$ consents. Replacing the
binary indicators with $v_i\in[0,1]$ produces $L$. The correspondence
$v_i=\mathbb 1[\operatorname{Consent}_i]$ is sufficient to make $L=q$ and hence to reproduce the
authorization event at every threshold. It is not necessary for event equivalence at one fixed
threshold. At $\theta=1$, the exact condition is
\begin{equation*}
\bigl[\,L=1\,\bigr]
\iff
\bigl[\,\operatorname{Consent}_i=1\ \text{for every }i\in S_d\,\bigr].
\end{equation*}
Under an agentwise coding $v_i=g_i(\operatorname{Consent}_i)$, this equivalence holds for every
consent profile if and only if $g_i(1)=1$ and $g_i(0)<1$ for each positive-stake agent. Thus a
non-consenter may receive any voice score below one without changing the unanimity event.
\end{observation}

\begin{proof}
Divide both sides of Equation~\eqref{eq:authorization} by the positive total stake $\sigma(d)$ to
obtain $q(d)\geq\theta$. If $v_i=\mathbb 1[\operatorname{Consent}_i]$ for all affected agents,
then $L=q$ and every threshold event agrees. At $\theta=1$, a positive-weighted average of values
in $[0,1]$ equals one if and only if every value equals one. Hence the displayed biconditional is
the necessary-and-sufficient event-level condition.
For an agentwise coding, the all-consenting profile requires $g_i(1)=1$ for every $i$, while the
profile in which only agent $i$ withholds consent requires $g_i(0)<1$; these conditions are also
sufficient. They do not require $g_i(0)=0$.
\end{proof}

The observation is modest but load-bearing. It prevents a continuous score from inheriting the
normative status of unanimity by resemblance alone. $L$ can be useful for comparing influence,
including when no binary authorization record exists. In that case it should be reported as
effective voice. Whether a graded voice threshold is normatively adequate is a separate premise.

\subsection{Threshold margins and classification uncertainty}

Authorization is often reported as a binary classification even though affected-set membership,
stakes, and consent records are uncertain. The threshold margin
\begin{equation}
m_\theta(d)=q(d)-\theta
\label{eq:threshold-margin}
\end{equation}
makes that uncertainty visible. A large positive margin is robust to small perturbations of the
declared weights; a margin near zero can change sign under minor measurement choices. The margin
does not soften the axiom. At $\theta=1$, any verified positive-stake non-consenter makes
$m_1<0$. It provides a diagnostic for data and calibration uncertainty around relaxed thresholds.

Suppose plausible stake specifications form a set $\mathcal S$ and affected-set rules form
$\mathcal B$. A robust classification can be defined by the interval
\begin{equation}
\mathcal M_\theta(d)=
\left[
\inf_{s\in\mathcal S,\,b\in\mathcal B}m_\theta(d;s,b),
\sup_{s\in\mathcal S,\,b\in\mathcal B}m_\theta(d;s,b)
\right].
\label{eq:threshold-interval}
\end{equation}
If the interval lies above zero, the decision satisfies the chosen relaxed rule across the declared
sensitivity set. If it lies below zero, it fails throughout. If it crosses zero, the classification
depends on unresolved measurement choices and should be reported as such. This sensitivity
exercise does not determine which choices are normatively correct; it shows whether the conclusion
is doing hidden work through them.

Missing authorization data require the same discipline. Coding non-response as consent maximizes
$q$; coding it as non-consent minimizes $q$. Where missingness could depend on coercion, access, or
stake, a missing-at-random analysis is not automatically credible. A study should report bounds
under explicit missing-data assumptions and distinguish absence of a recorded objection from a
recorded affirmative act. For the unanimity axiom, unknown authorization for a positive-stake agent
prevents a verified positive classification.

Threshold comparison also changes the estimand. Estimating outcomes just above and below a legal
quorum can identify consequences of the rule under suitable continuity assumptions, but it does
not identify the moral correctness of $\theta$. Nor does an empirically convenient threshold make
stake shares measured under different rules comparable: agenda setting and participation can
change endogenously with the threshold. The authorization layer should therefore report the rule,
margin, sensitivity set, and behavioral response separately.

Effective influence is not the same as a formal vote. Agenda control, vetoes, defaults, exit
options, delegation, and implementation discretion can make nominally equal votes unequally
effective. Conversely, observed policy proximity does not prove that an agent had voice: the
controller may have independently selected the same outcome. Authorization data and outcome data
therefore cannot substitute for a measure of influence.

\subsection{An authorization protocol}

Applying Equation~\eqref{eq:authorization} requires a protocol, not merely a threshold. First,
define the commitment. A continuing membership rule, a one-off allocation, and an irreversible
transfer are different decisions even when they concern the same resource. Second, define the
affected set using a prospective exposure criterion rather than observed complaint. Parties whose
harm is suppressed or delayed do not acquire zero stake because they are silent at measurement
time. Third, specify the authorization act and its expiry. A vote, signed delegation, opt-in, or
revocable instruction may qualify under different institutional rules; mere presence after the
decision does not.

Fourth, audit the option set. An agent choosing the least harmful of two imposed alternatives may
authorize that selection without authorizing the prior exclusion of other options. The operative
locus can therefore be nested: one procedure sets the menu, another selects within it, and a third
implements the result. Authorization at one locus does not transfer automatically to the others.
Fifth, audit revocation and exit. If a delegation cannot be withdrawn at reasonable cost, continued
control may persist without continuing authorization. Finally, report disagreement. A unanimity
failure is not missing data; it is the outcome of the normative test.

The stakes weights also need an interpretation. They can represent expected welfare loss over a
declared contrast, rights-relevant exposure, or another normatively defended quantity. They should
not combine benefits and harms in a way that makes an agent with a chance of catastrophic loss
appear equivalent to many agents with trivial gains unless that trade-off is intended. Absolute
local utility gradients are one operationalization for resource allocation, not a general theory
of moral importance. Distributional summaries and worst-case exposures should accompany an
aggregate whenever tail stakes matter.

Three edge cases illustrate the protocol. In a delegated decision, the representative is the
operative locus; the affected agents' earlier authorization of the delegation is relevant only
within its scope and duration. In a default decision, inaction still selects the default, so the
designer of the default and the rule governing non-response belong to the locus. In an emergency,
a lower threshold may be operationally justified by time pressure, but the record should state the
relaxation, the reason, the expiry, and any retrospective authorization or remedy. None of these
cases is clarified by calling every effective controller a consent-holder.

\subsection{A locus-by-locus authorization register}

Complex decisions should be audited as chains rather than assigned one controller label. A minimal
register has one row for each load-bearing locus: agenda formation, option-set construction,
terminal selection, implementation, monitoring, appeal, and revision. For each row the analyst
records the operative controller, the action controlled, the affected set, the stake calibration,
the qualifying authorization act, its scope and expiry, the applicable threshold, and the remedy
for refusal or breach. The register is a reporting device, not an additional outcome model.

This decomposition prevents authorization from transferring by proximity. Suppose affected agents
authorize a representative to choose among a declared set of allocations for one year. That act
can authorize the representative at the terminal-selection locus within the stated menu and time
window. It does not by itself authorize whoever constructed the menu, a later expansion of the
resource domain, an implementation agent that substitutes a different allocation, or indefinite
renewal. Transfer requires an explicit delegation relation whose object, duration, revocation rule,
and downstream powers are recorded. A formally valid link at one locus cannot fill a missing link
elsewhere.

The same register distinguishes three kinds of failure. A \emph{classification failure} occurs
when the declared rule is not met: for example, one positive-stake party refuses under the
unanimity axiom. A \emph{measurement failure} occurs when the evidence cannot establish the inputs:
the affected set is incomplete, stakes are incommensurable, or an apparent opt-in is not reliably
observed. A \emph{scope failure} occurs when a genuine authorization act is used outside its
declared object, time, or delegation depth. These statuses should not be pooled into a generic
``low consent'' score. They imply different corrections and different uncertainty.

Authorization is also time-indexed. A decision may be authorized when made but become
unauthorized after material changes to stakes, participants, information, or available options.
Conversely, retrospective ratification can authorize continued operation without making the prior
imposition authorized. A longitudinal study should therefore distinguish the authorization time,
the operative-control interval, and the outcome window. Measuring them at one date and applying
the result to another silently assumes stability in the very relations at issue.

For distributed or automated systems, the register should include counterfactual influence. A
formal veto that cannot be exercised before an automated deadline may contribute little effective
voice even though it appears in the rules. An algorithm that merely recommends has a different
locus from one whose output executes by default. The relevant question is which intervention would
change the selected commitment while the remaining procedure is held fixed. This counterfactual
does not establish consent, but it identifies the operative locus to which the authorization test
must be applied.

The appropriate summary is therefore a vector of locus-specific classifications, not an automatic
single score. A study may report that agenda setting is unverified, selection satisfies a named
supermajority relaxation, implementation remains within delegation, and appeal is unavailable.
That report is less rhetorically tidy than ``the system is legitimate,'' but it exposes where the
normative and empirical work actually occurs.

\subsection{Alignment, stakes, and information}

The model layer begins with three candidate coordinates. They are not terms in the authorization
predicate except insofar as stakes appear in both layers under a common calibration.

\begin{definition}[Objective alignment]
\label{def:alignment-reader}
For affected agent $i$, operative controller $h_d$, and a declared probability measure $\mu_d$ over
relevant states or allocation contexts, define
\begin{equation}
\alpha_i(d,t)=\operatorname{corr}_{\mu_d}\!\left(T_i,T_{h_d}\right)
=\frac{\operatorname{Cov}_{\mu_d}(T_i,T_{h_d})}
{\sqrt{\operatorname{Var}_{\mu_d}(T_i)\operatorname{Var}_{\mu_d}(T_{h_d})}}
\in[-1,1].
\label{eq:alignment-reader}
\end{equation}
The measure $\mu_d$ and nonzero finite variances are part of the definition. Constant objectives
require a distance-based alternative.
\end{definition}

Alignment concerns the direction of objectives over a specified support. It is not authorization.
A benevolent controller can be highly aligned without having been appointed by affected agents;
an authorized representative can make a locally misaligned decision. Correlation is also
distribution-relative. Changing the states on which rewards are evaluated can change both the
delivered treatment and the inference. The companion's target-support repair is an empirical
instance of this general point.

When a single aggregate is required, the stakes-weighted mean
\begin{equation}
\bar\alpha(d,t)=
\frac{\sum_{i\in S_d}s_i(d)\alpha_i(d,t)}{\sum_{i\in S_d}s_i(d)}
\label{eq:aggregate-alignment-reader}
\end{equation}
is transparent, but it discards heterogeneity. Because $x\mapsto(1+x)^{-1}$ is strictly convex on
$(-1,1]$, plugging $\bar\alpha$ into the canonical ratio understates the corresponding additive
index whenever individual alignments vary. This aggregation issue is mathematical, not an
argument that either index is empirically valid.

Stakes require stronger discipline than a label such as ``high'' or ``low.'' Suppose resource $r$
is changed by an increment $\Delta a_r>0$ and $U_{0,S}>0$ is a common utility-loss reference at
scale $S$. A local exposure can be represented by
\begin{equation}
\sigma_i^{\mathrm{raw}}(r)
=\left|\frac{\partial U_i}{\partial a_r}\right|\Delta a_r,
\qquad
\widetilde\sigma_i(r)=\frac{\sigma_i^{\mathrm{raw}}(r)}{U_{0,S}}.
\label{eq:stake-calibration}
\end{equation}
The increment removes dependence on the physical unit chosen for $a_r$; the reference makes the
quantity dimensionless. Aggregating across agents or resources is meaningful only if their
utility-loss scales and the reference are declared commensurable. Multiplying every reward by a
parameter labeled ``stake'' is not empirical evidence that calibrated stakes govern coordination: it mechanically
rescales the loss surface.

Information deficit must specify a channel and a target variable. For a discrete or discretized
preference variable $P_i$ with $0<H(P_i)<\infty$ and signal $Z_{i\to h_d}$ observed by the controller,
one candidate is
\begin{equation}
\varepsilon_{i h_d}
=\frac{H(P_i\mid Z_{i\to h_d})}{H(P_i)}
=1-\frac{I(P_i;Z_{i\to h_d})}{H(P_i)}\in[0,1].
\label{eq:information-deficit}
\end{equation}
This is residual uncertainty under a declared distribution. It is undefined when $H(P_i)=0$ and
does not measure whether the controller favors or opposes the preference. Mutual information and
transfer entropy are sign-blind: deterministic inversion can be as predictable as faithful
response. A directional objective measure and an information-channel measure therefore cannot be
collapsed merely because both are functions of observed behavior.

\subsection{The friction ansatz}

For a homogeneous or aggregate representation, the canonical model index is
\begin{definition}[Canonical friction index]
\label{def:friction-reader}
For $\sigma\geq0$, $\varepsilon\in[0,1]$, and $\alpha\in(-1,1]$, let
\begin{equation}
F(\alpha,\sigma,\varepsilon)
=\sigma\frac{1+\varepsilon}{1+\alpha}.
\label{eq:friction-reader}
\end{equation}
$F$ is a dimensionless index only when $\sigma$ has been normalized as in
Equation~\eqref{eq:stake-calibration}. It is neither the authorization predicate nor an observed
coordination outcome.
\end{definition}

Equation~\eqref{eq:friction-reader} is the unit-normalized member of a restricted rational--affine
class. Its form was chosen to encode higher modeled friction with stakes and information deficit,
lower modeled friction with cooperative alignment, a positive stake-proportional floor at perfect
alignment, and a singular adversarial boundary. These features are desiderata, not data. The
derivatives
\begin{equation}
\frac{\partial F}{\partial\sigma}=\frac{1+\varepsilon}{1+\alpha}>0,
\qquad
\frac{\partial F}{\partial\varepsilon}=\frac{\sigma}{1+\alpha}\geq0,
\qquad
\frac{\partial F}{\partial\alpha}
=-\frac{\sigma(1+\varepsilon)}{(1+\alpha)^2}\leq0
\label{eq:comparative-statics}
\end{equation}
are algebraic consequences of the selected function; the latter two inequalities are strict when
$\sigma>0$. They cannot validate the variables, their
composition, or the outcome link. In particular, the cross-partial
\begin{equation}
\frac{\partial^2F}{\partial\alpha\,\partial\varepsilon}
=-\frac{\sigma}{(1+\alpha)^2}\leq0
\label{eq:cross-partial}
\end{equation}
is strictly negative when $\sigma>0$ and is a testable qualitative implication: the marginal index penalty from information loss shrinks
as alignment rises. In the companion's reduced feasible-centred frozen-policy crossing, the
exploratory target-correlation--observation-noise interaction for the held-out reward gap has the
opposite sign; this is a proxy-level contradiction, not a direct test of general alignment or
normalized conditional entropy.

The positive floor $F=\sigma/2$ at $(\alpha,\varepsilon)=(1,0)$ follows from normalization, not
from an observed irreducible delegation cost. The divergence as $\alpha\to-1$ follows from the
denominator, not from evidence of an empirical singularity. The symmetric alternative
$F^{(2)}=\sigma(1+\varepsilon)/(1+\alpha^2)$ remains a formal example under different boundary
desiderata, but its original empirical motivation was a sign-blind artifact and is withdrawn. The
companion's exploratory target-coordinate proxy comparison supports neither composite as a
single predictor of its specified reward-gap outcome.

An additive stakeholder form
\begin{equation}
F_{\mathrm{add}}=\sum_{i\in S_d}s_i\frac{1+\varepsilon_i}{1+\alpha_i}
\label{eq:additive-friction}
\end{equation}
is not generally equal to applying Equation~\eqref{eq:friction-reader} to aggregate coordinates.
With common entropy, Jensen's inequality makes the aggregate-alignment form a lower bound. This is
one reason an empirical protocol must state its aggregation rule before observing outcomes.
Alternative exponential, geometric, or divergence-based functions can share some directional
properties while differing at boundaries and in interactions. The long technical record
\TSpointer{TS-FORM} gives the full desiderata, agency-theoretic motivation, and sensitivity
catalogue. No motivation there upgrades the ratio from ansatz to derivation.

\subsection{Index, authorization, legitimacy, and outcome}

The framework's most important negative definition is that $F\neq Y$. Let $Y_d$ be a prespecified
observed outcome in domain $d$: reward gap, deadlock time, communication cost, exit, noncompliance,
or another construct with its own measurement model. The empirical claim takes a form such as
\begin{equation}
Y_d=g\!\left(F_d,X_d\right)+u_d,
\label{eq:index-outcome}
\end{equation}
where $X_d$ contains confounders or design factors and $g$ is declared before comparison. Failure
of Equation~\eqref{eq:index-outcome} rejects that index--outcome specification. It does not refute
the unanimity principle. Conversely, satisfaction of the authorization predicate does not imply
low $Y$, and a low $Y$ cannot establish authorization.

The effective-voice score $L$ occupies another layer. Under the declared normative bridge it can
be interpreted as a legitimacy measure; descriptively, it summarizes stakes-weighted influence.
$F$ summarizes measured objective correlation, stake, and information under an ansatz. They can
move independently. An agent may have influence but pursue an objective poorly correlated with
other affected agents. A controller may be aligned by coincidence but lack authorization. A
decision can be unanimously authorized and still carry high implementation cost because its
stakes are large. These are not anomalies. They are cases the separation is designed to represent.

This architecture also narrows the normative claim. Scanlonian contractualism asks which
principles no one could reasonably reject \citep{scanlon1998we}; the Axiom uses actual authorization
and stakes but does not supply a complete account of coercion, competence, option quality, or
reasonable rejection. Social-choice theory shows why aggregation procedures confront hard
constraints; it does not decide which agents enter $S_d$ or whether a measured correlation counts
as voice. The framework contributes a transparent authorization rule and an empirical contract,
not a complete moral theory.

\subsection{What stakes weighting does and does not settle}

Weighting authorization by stakes is intended to prevent a numerically large low-exposure group
from automatically overriding a smaller group that bears the decision's consequences. It extends
the all-affected principle from binary standing to magnitude. Yet it creates its own normative and
measurement burdens. Stakes can be multidimensional, incomparable, uncertain, endogenous to the
decision, or strategically reported. A monetary proxy can miss dignity, rights, or irreversible
harm. A subjective utility scale can be transformed without preserving interpersonal ratios. A
computational reward may be an engineering signal rather than an agent's welfare.

That proportional rationale applies to relaxed thresholds and continuous effective-voice
summaries. At the literal unanimity boundary $\theta=1$, every positive-stake agent has a veto:
stake magnitudes determine affected-set standing through positive versus zero exposure but do not
change the authorization classification once an agent is included.

The framework therefore uses calibrated stakes as declared model inputs, not natural quantities.
A protocol should show how $s_i$ changes under plausible scale choices, whether the authorization
classification is robust to those changes, and which exposures were excluded. When stakes are not
commensurable, vector-valued reporting or separate affected classes may be more honest than a
forced scalar. The axiom says that stakes matter for standing; it does not guarantee that every
stake can be measured on one cardinal scale.

Nor does weighting replace side constraints. If a protected right is treated as non-tradeable,
that restriction should enter the admissible decision set before aggregation. Otherwise a large
sum of small benefits can numerically authorize an infringement the normative theory meant to
exclude. AOC is compatible with such constraints but does not derive them. Its threshold rule is a
component of an authorization account, not the whole account.

The same caution applies to negative and positive stakes. Equation~\eqref{eq:authorization} uses
positive exposure magnitudes and a consent indicator; it does not net an agent's possible benefit
against another agent's harm before determining who is affected. Net welfare can be evaluated in a
separate objective. Keeping the affected set broad prevents a favorable aggregate forecast from
erasing a party whose authorization is still normatively required.

These qualifications make the principle harder to operationalize, but they prevent an apparent
precision from doing normative work. The empirical contract can compare institutions under a
declared stake rule. It cannot settle all disputes about that rule with the same data used to fit a
coordination outcome.

\subsection{Concentrated, distributed, and encoded loci}

The form of a decision locus changes how authorization and voice should be measured. In a
concentrated locus, one agent or office has final authority. Formal appointment identifies one
possible authorization chain, while agenda access, review, removal, and scope determine whether the
chain remains valid. Measuring only whether the controller's decisions match stakeholder
preferences would confuse performance with authorization.

In a distributed locus, influence may be unequal even when every participant has an action. Voting
weights, proposal rights, quorum rules, sequential moves, network centrality, and information
access can determine effective voice. Equation~\eqref{eq:legitimacy} therefore uses $v_i$ rather
than a binary participation flag. Estimating $v_i$ can require interventions or counterfactual
influence analysis: how often would the selected outcome change if agent $i$'s message or action
changed while other inputs were held fixed? A raw message count is not enough.

An encoded locus places authority in a protocol, model, or smart contract. The code executes a
prior allocation of decision rights. Relevant authorization questions include who selected the
code, who can modify it, which agents could meaningfully refuse it, and what happens when the
encoded rule encounters an unanticipated case. Immutability can constrain later controllers while
also making revocation costly. It does not remove the decision locus; it distributes it across
design, deployment, update, and execution stages.

Markets and randomization are delegated loci. A market outcome follows from rules governing
property, bids, matching, and enforcement. A lottery follows from the decision to use a random
device and from the choice of its option set and probabilities. Saying ``the market decided'' or
``chance decided'' can describe the terminal selection while hiding the meta-level controller.
Authorization must be evaluated at the level where the rule was adopted and maintained.

Hybrid systems combine these forms. A learned policy may recommend an action, a human may hold a
formal veto, and an automated default may execute unless the veto occurs within a short window.
The operative locus depends on actual influence: a veto that cannot be exercised with available
information may provide little effective voice. Mapping the chain is empirical work. The Axiom
does not assign consent to the most visible component by default.

These distinctions also guide outcomes. Concentrated control may reduce decision time while
lowering voice; distributed control may increase deliberation cost while making broader
authorization feasible if qualifying acts occur; encoded control may be predictable but brittle.
Architecture alone is not an authorization act. There is no reason to expect one scalar outcome to
rank every architecture. Layered reporting permits genuine trade-offs without treating efficiency
as legitimacy or delay as evidence against consent.

\section{A conditional persistence interface to ROM}
\label{sec:rom-interface}

The Axiom of Consent does not require an evolutionary model. The Replicator--Optimization
Mechanism (ROM) is an optional persistence interface: it asks what follows if coordination
configurations reproduce, transform, or disappear according to a declared effective-fitness law
\citep{farzulla2026rom}. AOC owns the authorization variables passed into that interface. ROM owns
the persistence dynamics, kernel formalism, scale transformations, and identification results. This
section states only the contract between them.

\subsection{Configurations and the kernel triple}

Fix a scale $S$ and finite type set $T_S$. A type $\tau\in T_S$ is a coordination configuration,
such as an allocation of decision rights or a policy architecture. Let $p_t(\tau)$ denote its
population share. ROM assigns each configuration a triple
\begin{equation}
\bigl(\rho_S(\tau,G,p),\;w_S(\tau),\;M_S(\tau\to\tau')\bigr),
\label{eq:rom-triple-reader}
\end{equation}
where $\rho_S$ is a nonnegative bounded persistence or survival factor, $w_S$ is a nonnegative scale-relative
weight, and $M_S$ is a row-stochastic transformation kernel. Network structure $G$ and population
state $p$ may enter $\rho_S$ in the general model. The AOC instantiation uses the effective-voice
score interpreted through a declared legitimacy bridge, together with a normalized friction
index, to propose one survival factor; it does not identify the other ROM
components from the authorization rule.

The dynamics are
\begin{equation}
\frac{dp_t(\tau)}{dt}
=\sum_{\tau'\in T_S}p_t(\tau')w_S(\tau')
\rho_S(\tau',G,p_t)M_S(\tau'\to\tau)
-p_t(\tau)\bar\phi_t,
\label{eq:rom-reader}
\end{equation}
with
\begin{equation}
\bar\phi_t=
\sum_{\tau\in T_S}p_t(\tau)w_S(\tau)\rho_S(\tau,G,p_t).
\label{eq:mean-fitness-reader}
\end{equation}
The subtraction term preserves total mass. When $M_S$ is the identity and $w_S=1$, the equation
reduces to a replicator form. Transformation, learning, institutional replacement, or mutation can
be represented through $M_S$, but the semantic interpretation must be specified in the domain.

Two cautions follow immediately. First, the triple is scale-relative. An individual policy, team
architecture, and institution are not automatically comparable types. Second, the decomposition
of effective fitness into $w_S$ and $\rho_S$ is not identified by population trajectories alone:
their product drives Equation~\eqref{eq:rom-reader}. ROM develops the conditions under which
different decompositions or coarse-grainings can be compared. AOC does not reproduce those
results.

\subsection{The consent-friction bridge}

Let $F(\tau)$ be the dimensionless candidate index for configuration $\tau$, constructed from
stakes normalized by the declared reference in Equation~\eqref{eq:stake-calibration}. Let
$L(\tau)\in[0,1]$ be the stake-weighted effective-voice score from
Equation~\eqref{eq:legitimacy}, interpreted as legitimacy only under the stated normative bridge.
The consent-friction
instantiation proposes
\begin{equation}
\rho_S^{\mathrm{AOC}}(\tau)
=\frac{L(\tau)}{1+F(\tau)}.
\label{eq:survival-bridge}
\end{equation}
This equation is a bridge law. It says that, within the model, a larger bridged effective-voice
measure raises
the survival factor and a larger index lowers it. Neither direction follows from the
structural existence of a decision locus or from the unanimity axiom. The law requires independent
testing against persistence outcomes.

Normalization is load-bearing. The stake reference used to construct $F$ must be fixed within a
comparison and justified across scales. An optional rebaselining $\widetilde F=F/F_{0,S}$ would
define a different survival map unless $F_{0,S}=1$; it must therefore be declared and tested as a
distinct model, not inserted silently into Equation~\eqref{eq:survival-bridge}. The map is also not
the only admissible bridge: $e^{-F}$ or another decreasing function could encode the same
qualitative premise while implying different dynamics. Empirical adequacy of the canonical
friction ratio cannot validate Equation~\eqref{eq:survival-bridge}, and failure of the ratio does
not by itself reject every possible persistence model.

The bridge keeps authorization and selection distinct in a useful way. The predicate
$\operatorname{Authorized}_\theta$ is binary and normative. $L$ is a continuous measured
relaxation. $F$ is a model index. $\rho_S$ is a posited persistence factor. $p_t$ is a population
state. A chain from authorization to long-run prevalence therefore contains at least four arrows:
authorization measurement, index construction, survival mapping, and population dynamics. Any
arrow can fail without logical contradiction.

\subsection{Exactly what the stationary result permits}

For general frequency-dependent replicator--mutator systems, convergence is not guaranteed;
cycles and more complicated dynamics are possible \citep{hofbauer1998evolutionary,galla2013complex}.
ROM proves a restricted stationary-composition result for the autonomous continuous-time flow;
this reader uses only its conditions and consequence. Let the fixed effective-fitness vector be
$\phi_S(\tau)=w_S(\tau)\rho_S(\tau)>0$. A unique normalized positive Perron composition and
convergence of the continuous-time flow follow under the following conditions:
\begin{enumerate}[noitemsep]
\item $T_S$ is finite;
\item the interaction structure and all fitness components are static;
\item $w_S(\tau)>0$ and survival is positive and bounded;
\item survival is density-independent, so $\rho_S(\tau,G,p)=\rho_S(\tau)$ for fixed $G$; and
\item the transformation kernel is irreducible.
\end{enumerate}
Zero-survival types permitted by the general interface fall outside this strict-positivity theorem.
Under those assumptions the fixed nonnegative operator
$Q_S=\operatorname{diag}(\phi_S)M_S$ is irreducible, and Perron--Frobenius theory provides a unique
normalized positive left eigenvector. ROM's continuous-time proof shows that the normalized flow
converges to that composition. Aperiodicity is not required for either conclusion. If $Q_S$ is
instead used in a normalized discrete-time power iteration, aperiodicity makes it primitive and is
a sufficient additional condition for convergence rather than peripheral cycling. None of these
results is universal convergence of adaptive institutions, changing networks, or
frequency-dependent multi-agent learning.

The operator proof and stationary algebra are not reproduced here. The load-bearing boundary is
that state-dependent fitness, changing networks, zero-fitness types, or disconnected transition
classes remove one or more premises of the restricted result. Periodicity does not remove the
unique Perron composition or continuous-time convergence, but it can defeat normalized
discrete-time power convergence. Even when uniqueness survives, the componentwise ordering
established here is guaranteed under the exact uniform-residual kernel used in the
technical theorem. Under a general kernel, stationary mass also depends on inflow from other
types; under type-specific mixing, faster transformation away from a type can offset higher own
fitness. Comparisons across information-deficit levels are therefore not settled by
$L/(1+F)$ alone. \TSpointer{TS-ROM} gives stable destinations for the matrix construction,
stationary equation, and ranking derivation in the ROM companion \citep{farzulla2026rom}.

The practical conclusion is deliberately conditional:
\begin{quote}
If configurations evolve under a finite, static, density-independent ROM; if the survival factor
is stipulated as $L/(1+F)$; and if the specified residual kernel supplies the ranking condition,
then configurations with larger effective fitness receive more stationary mass. At equal weights
and fixed entropy, larger $L$ and smaller $F$ raise that effective fitness.
\end{quote}
Removing the antecedents removes the conclusion. The result does not show that consent-respecting
arrangements must prevail, that unauthorized systems are unstable, or that persistence certifies
legitimacy.

\subsection{How the interface could be tested}

Testing Equation~\eqref{eq:survival-bridge} requires observations at the configuration level. A
type must be defined before its frequency is measured; births, transitions, mergers, and exits must
be assigned consistently; $w_S$, $L$, and $F$ must be measured without using the same duration
outcome; and competing risks must be represented. A regression of duration on a label such as
``consensual'' would not test the proposed decomposition.

One design would estimate primitive-level quantities at time $t$, construct candidate survival
maps, and compare their out-of-sample prediction of transitions over $[t,t+\Delta]$. The canonical
bridge should be compared with at least effective-voice-only, index-only, additive, and flexible
interaction alternatives. If a lower-index configuration persists only because exit is blocked,
variables measuring coercion or transition opportunity belong in the model. If type definitions
change during the interval, a static ROM is inappropriate.

Identification also requires variation. If all observed configurations have similar $L$, the
effective-voice channel cannot be estimated. If $L$ and $F$ are mechanically related through shared
inputs, their separate coefficients are not identified without an intervention or credible
restriction. If the transformation kernel is unknown, changes in type frequency can reflect
inflow rather than own survival. Panel data can help, but repeated observations do not remove
time-varying confounding or misclassification.

The empirical target should match the theorem. A test of one-period transition probability is not
a test of convergence to a unique stationary composition. Evidence of an approximately stable
frequency over a short window does not establish stationarity or uniqueness. Conversely, failure
to observe convergence in a changing, frequency-dependent system does not refute the restricted
Perron--Frobenius result; it shows that the result does not govern that system. The hypotheses must
be checked rather than assumed from notation.

The transformation kernel merits its own diagnostics. Row-stochasticity is an accounting
condition. Irreducibility says every type can eventually reach every other type, which may be false
when legal, technological, or biological constraints create disconnected classes. Aperiodicity
excludes peripheral cycles in the normalized discrete-time power iteration; it is not needed for
the unique Perron composition or convergence of the continuous-time normalized flow. Uniform residual exploration is especially restrictive: it gives
the componentwise ranking but is rarely a natural institutional transition law. Reporting the
stationary ranking without reporting the kernel would therefore omit a load-bearing assumption.

Finally, the interface needs negative controls. A shuffled effective-voice score should not predict
transitions. An outcome measured after the transition should not be used as a baseline covariate.
Alternative type partitions should reveal whether apparent persistence is an artifact of
aggregation. Where possible, externally imposed changes in decision rights or information can
separate survival, weight, and mutation channels. These requirements make ROM an empirical
programme rather than a metaphor for institutional evolution.

\subsection{Scale and transformation as interface limits}

The scale index $S$ prevents an easy but invalid inference from local alignment to system-level
authorization. A controller can align with a team while the team's objective conflicts with
external stakeholders. Conversely, local disagreement can coexist with an authorized global
procedure. Aggregating types, stakes, or transition probabilities must preserve the relevant
dynamics if a macro-level ROM description is claimed.

ROM formalizes this problem through exact lumpability. Informally, a partition of fine types is
lumpable only when every fine type within a macro-block sends the same total effective-fitness-
weighted flow to every other macro-block. Under that condition the projected process admits a
closed ROM equation. Weaker similarity is not enough: hidden fine-state dependence can produce
memory after projection. Moreover, exact lumpability identifies a block-level effective-fitness
product; separately recovering coarse survival and weight requires stronger homogeneity
conditions. The proof, identification transformation, and scale-transition examples belong to the
ROM companion \citep{farzulla2026rom}; \TSpointer{TS-ROM} gives their stable reader and supplement
destinations.

The interface rule for AOC is consequently simple. Every empirical claim names the scale at which
the affected set, stake reference, alignment measure, information channel, outcome, and type
distribution are defined. A macro-level authorization claim cannot be inferred from a micro-level
index without a substantive aggregation argument. A repeated symbol across scales is not
measurement invariance.

\subsection{Why ownership and belief transfer are not part of the core claim}

The long technical record includes an extension in which longer operative-control tenure raises a
latent ownership-perception score and changes transition probabilities. Under a positive incumbent
self-loop and fixed alternative targets, the normalized exit probability decreases with tenure.
That proposition is mathematically useful, but it is an added kernel specification. It neither
follows from the authorization axiom nor identifies psychological ownership from observed duration.
Negative duration dependence can arise through selection, repression, learning, or unmeasured
heterogeneity.

For the same reason, information deficit is not automatically the ROM mutation rate. A monotone
map from measured $\varepsilon$ to a mixing parameter must be chosen and calibrated. The empirical
finding that observation noise affects a coordination outcome does not prove that information loss
causes institutional mutation. The reader paper therefore leaves belief transfer,
ownership-sensitive transformation, and their proofs to ROM; \TSpointer{TS-ROM} supplies exact
navigation. Their status is extension, not support for the core authorization architecture.

\subsection{The normative firewall}

Evolutionary explanations can be illuminating without being justificatory. A coercive arrangement
may persist because exit is blocked; a legitimate arrangement may disappear after an external
shock. Equation~\eqref{eq:survival-bridge} deliberately puts $L$ into modeled survival, but doing so
is a substantive bridge principle: it hypothesizes that effective voice contributes to persistence
under stated conditions. Evidence that the hypothesis predicts duration would support that
descriptive link. It would not prove that persistence is good, that every source of persistence is
legitimate, or that the measured voice score captures genuine authorization.

This firewall is the main reason to retain ROM as an interface rather than as AOC's backbone. The
authorization principle can be debated on normative grounds even if the persistence model fails.
ROM can be studied as a formal dynamics without accepting the axiom. Empirical work can reject the
canonical index while still asking whether better measures of voice or objective conflict predict
coordination. The three projects meet at declared mappings, not at an assertion of conceptual
identity.

The division also prevents duplicate claim ownership across the paper family. ROM's equation,
kernel triple, exact lumpability results, reversibility and probability-current distinctions, and
identification transformations should be evaluated in the ROM paper. The MARL paper owns the
control-battery evidence and rejection of the composite. AOC owns the authorization rule,
decision-locus distinction, legitimacy interpretation, and measurement layers. Each paper can
state the status of an interface, but none should reproduce a neighbor's result as if it were an
independent contribution.

\section{Primary instantiation: multi-agent resource allocation}
\label{sec:multiagent}

Resource allocation is the principal application because it makes control concrete. Let
$\mathcal A=\{1,\ldots,n\}$ be agents, $\mathcal R=\{1,\ldots,m\}$ resources, and
$x_t\in\mathcal X$ the resource state. Each agent observes $o_{i,t}$, selects an action
$a_{i,t}\in\mathcal U_i$, and receives reward $R_i(x_t,a_t)$. A transition map
$P(x_{t+1}\mid x_t,a_t)$ determines how joint actions change resources. This is not yet a consent
model. It becomes an AOC instantiation only after the operative locus, affected set, authorization
rule, calibrated stakes, objective comparison, information channel, and outcome are declared.

\subsection{Control and authorization in an MDP}

\begin{definition}[Operative resource controller]
\label{def:resource-controller-reader}
For a resource-control configuration $\Gamma$, agent or procedure
$h=\operatorname{ctrl}_{\Gamma}(r,t)$ is the operative controller of resource $r$ at time $t$ if its
action, veto, aggregation rule, or access permission determines the admissible allocation change
for $r$. If several actions are aggregated, the aggregation mechanism is itself part of the
decision locus. When $t$ is fixed or the configuration is time-invariant, we write
$\operatorname{ctrl}_{\Gamma}(r)$.
\end{definition}

This definition handles centralized and decentralized systems. A scheduler that alone allocates a
resource is a concentrated locus. A shared-pool transition $x_{r,t+1}=g(x_{r,t},\sum_i a_{ir,t})$
creates a distributed operative locus: each agent contributes to the joint move under the encoded
aggregation rule. Independent copies $x_{ir,t+1}=g(x_{ir,t},a_{ir,t})$ instead create separate
local loci. A team reward does not by itself make physical control shared, and a shared state does
not by itself make authorization equal.

For decision $d_{r,t}$, let $S_{r,t}$ contain agents whose utility changes over the declared
allocation contrast. Equation~\eqref{eq:authorization} then evaluates whether the operative
controller was authorized. Three observations prevent shortcuts:
\begin{enumerate}[noitemsep]
\item being coded as an agent does not guarantee inclusion in the affected set;
\item having an action does not imply effective voice if the transition or aggregator nullifies it;
and
\item receiving a favorable reward does not establish an authorization act.
\end{enumerate}
An experimental MARL environment can test coordination mechanics without implementing normative
consent. In the companion, ``alignment'' refers to a signed relationship among reward targets. It
does not mean the policies authorized each other.

\subsection{A calibrated model index}

For resource $r$, choose an allocation increment $\Delta a_r$ and common reference $U_{0,S}$.
Agent $j$'s calibrated exposure $\widetilde\sigma_j(r)$ is defined by
Equation~\eqref{eq:stake-calibration}. For controller $i=\operatorname{ctrl}_{\Gamma}(r)$ and
affected agent $j$, define
allocation alignment over a declared distribution $\mu_r$ by
\begin{equation}
\alpha_{ij}(r)=
\operatorname{corr}_{\mu_r}
\left(\frac{\partial U_i}{\partial a_r},
      \frac{\partial U_j}{\partial a_r}\right),
\label{eq:allocation-alignment}
\end{equation}
provided both series have finite nonzero variance. Let $\varepsilon_{ij}(r)$ measure the
controller's residual uncertainty about $j$'s relevant preference signal. The pairwise candidate
index is
\begin{equation}
f_j(r)=\widetilde\sigma_j(r)
\frac{1+\varepsilon_{ij}(r)}{1+\alpha_{ij}(r)}.
\label{eq:resource-index}
\end{equation}
When every term uses the same calibration, a system index can be written
\begin{equation}
f_{\mathrm{system}}=
\sum_{r\in\mathcal R}\sum_{j\neq \operatorname{ctrl}_{\Gamma}(r)}f_j(r).
\label{eq:system-index}
\end{equation}
The sum is an index, not a reward gap. Its additivity assumes that the declared units and
aggregation rule remain meaningful across resources. Interactions among resources can make this
assumption false even when each local measurement is well defined.

When the denominator below is positive, the corresponding effective-voice score is
\begin{equation}
L(\Gamma)=
\frac{\sum_{r\in\mathcal R}\sum_{j\in\mathcal A}
\widetilde\sigma_j(r)v_j(r)}
{\sum_{r\in\mathcal R}\sum_{j\in\mathcal A}
\widetilde\sigma_j(r)}.
\label{eq:resource-legitimacy}
\end{equation}
Again, $L$ and $f_{\mathrm{system}}$ are distinct. $L$ asks how stakes track influence; $f$ combines
exposure, objective association, and information under the ansatz. A design can score well on one
and poorly on the other.

\subsection{Allocation objectives are normative choices}

Standard allocation design might maximize aggregate utility,
\begin{equation}
\Gamma^{U}\in\arg\max_{\Gamma}\sum_jU_j(\Gamma).
\label{eq:utility-allocation}
\end{equation}
If one instead takes the model index as the design objective, an index-minimizing allocation is
\begin{equation}
\Gamma^{F}\in\arg\min_{\Gamma}
\sum_r\sum_{j\neq \operatorname{ctrl}_{\Gamma}(r)}
\widetilde\sigma_j(r)
\frac{1+\varepsilon_{\operatorname{ctrl}_{\Gamma}(r),j}(r)}
{1+\alpha_{\operatorname{ctrl}_{\Gamma}(r),j}(r)}.
\label{eq:index-allocation}
\end{equation}
The two allocations can differ. That difference does not make $\Gamma^F$ universally better. It shows
that choosing an objective carries normative content. Aggregate utility, stake-weighted voice,
worst-case protection, equality, and predicted coordination cost answer different questions. The
Axiom constrains authorization; it does not prove that minimizing the rejected canonical index is
the correct welfare criterion.

This boundary is particularly important for high-stake minorities. A utilitarian objective can
select an allocation with large aggregate benefit and concentrated harm. Stake weighting makes the
exposure visible, but a numeric weight cannot by itself settle rights, side constraints, or
compensation. Similarly, unanimity may make some collective decisions impossible; that practical
cost is an objection to address, not permission to describe a lower threshold as the axiom.

\subsection{Reward alignment in MARL}

For agents with rewards $R_i(s,a)$ and $R_j(s,a)$, a signed alignment coordinate is
\begin{equation}
\alpha_{ij}=
\frac{\E_{\mu}[R_iR_j]-\E_{\mu}[R_i]\E_{\mu}[R_j]}
{\sqrt{\operatorname{Var}_{\mu}(R_i)
       \operatorname{Var}_{\mu}(R_j)}}.
\label{eq:marl-alignment-reader}
\end{equation}
The sampling distribution $\mu$ over state--action pairs is part of the estimand. A generator that
uses a nonnegative mixture weight can be monotone in a nominal parameter while never producing
negative realized correlation. A target generator can also request negative latent correlation
while clipping many desired states to the same environmental boundary. Both failures occurred in
the lineage of the companion experiment; both change what ``opposition'' means.

Equation~\eqref{eq:marl-alignment-reader} is a general candidate operationalization, not the
implemented treatment in the companion. The companion manipulates signed cross-agent correlation
$\rho$ among target-vector coordinates and verifies that target sampler. It does not estimate the
correlation of full reward functions over a declared state--action distribution $\mu$. In the
quadratic environment, target correlation is a designed source of reward-objective association,
but equality with Equation~\eqref{eq:marl-alignment-reader} would require a separate bridge that
accounts for the state distribution, weights, clipping, and policy visitation. We therefore use
``target alignment'' for the experimental treatment and reserve ``reward-function alignment'' for
the more general estimand above.

For $n$ agents with equicorrelated targets, the admissible lower correlation is
$-1/(n-1)$. Thus four-agent opposition is bounded below by $-1/3$; the pairwise endpoint $-1$ is
not geometrically available. Any argument built on the singularity at $\alpha=-1$ is therefore
outside that design. More broadly, a symbolic coordinate's range must be compatible with the
joint distribution and the environment's feasible state space.

The companion's shared-pool environment lets all agents move the same state coordinates. Rewards
are quadratic losses around heterogeneous targets. This design makes alignment consequential
through physical contention: jointly moving a common state can satisfy aligned targets more
easily than dispersed targets. In a fully separable control, each agent moves a private copy and
its marginal problem is unchanged by cross-agent target correlation. For independent Q-learning,
the expected alignment slope is therefore structurally zero. For value decomposition, the
learner still receives an additive team objective, so a finite-sample non-detection is not an
equivalence result even though the optimal control problem factorizes.

Partial sharing supplies the more informative intervention. Of $m$ resource coordinates, let
$k$ use the shared transition and $m-k$ use private transitions, with shared fraction $w=k/m$.
Agents still interact through the partially shared environment and, under VDN, the team reward.
Pairing the same latent target problem, weights, initialization, and evaluation starts across $w$
turns the shared fraction into a within-problem intervention. It changes physical contention and
may also change credit assignment, so the supported interpretation is modulation by shared
coordinates, not identification of a unique psychological or algorithmic mechanism.

\subsection{Worked example: one shared coordinate}

Consider $n$ agents acting on one bounded shared state $x\in[0,B]$. Agent $i$ has target $\tau_i$
and positive loss weight $c_i$, with reward
\begin{equation}
R_i(x)=-c_i(x-\tau_i)^2.
\label{eq:quadratic-reward-example}
\end{equation}
The average joint reward can be written
\begin{equation}
\frac1n\sum_iR_i(x)
=-\bar c\,(x-\bar\tau_c)^2-V_c,
\label{eq:quadratic-decomposition}
\end{equation}
where
\begin{equation}
\bar c=\frac1n\sum_i c_i,
\qquad
\bar\tau_c=\frac{\sum_i c_i\tau_i}{\sum_i c_i},
\qquad
V_c=\frac1n\sum_i c_i(\tau_i-\bar\tau_c)^2.
\label{eq:weighted-target-example}
\end{equation}
The continuous joint optimum is $x^*=\operatorname{clip}(\bar\tau_c,0,B)$. The term $V_c$ is an
irreducible dispersion cost: no single shared state can satisfy heterogeneous targets at once.
Increasing positive target association tends to reduce dispersion across problems drawn from a
common signed design, but the relationship depends on the target distribution and the state box.
This is a physical achievability channel, not automatically a learning channel.

The example clarifies three artifacts. First, multiplying every $c_i$ by $\sigma$ multiplies the
reward loss by $\sigma$ without changing $x^*$, target dispersion, transition structure, or the
optimal policy. It can nevertheless change Q-value and loss scales and therefore neural
optimization dynamics. A raw reward-scale main effect is still imposed algebraically by units. Second, if targets are drawn around zero while
$x\in[0,B]$, negative $\tau_i$ values share the same clipped optimum at the lower boundary. Two
opposed latent targets can become less opposed after clipping, while cooperative positive targets
retain more of their requested relationship. Third, comparing a learned trajectory with the
single-state $x^*$ treats travel from the reset as if the controller could teleport. A finite-
horizon dynamic benchmark is needed.

Now give each agent actions $a_{i,t}\in\{-1,0,+1\}$ and define
\begin{equation}
x_{t+1}=\operatorname{clip}\left(x_t+\sum_i a_{i,t},0,B\right).
\label{eq:shared-transition-example}
\end{equation}
Agents contend because every action changes the same $x$. Even if a joint controller can choose the
aggregate move efficiently, decentralized policies must coordinate their contributions. Target
alignment can affect both the location of $x^*$ and whether individually preferred moves point in
the same direction. A policy gap can therefore combine feasibility, travel, decentralized
information, exploration, and learned coordination.

In the separable version, agent $i$ has private state $x_i$ and transition
\begin{equation}
x_{i,t+1}=\operatorname{clip}(x_{i,t}+a_{i,t},0,B).
\label{eq:private-transition-example}
\end{equation}
Each optimum is $\operatorname{clip}(\tau_i,0,B)$. If the marginal distribution of $\tau_i$ is
held fixed while only cross-agent correlation changes, an independent learner's expected private
problem is invariant to that correlation. The negative-control slope is therefore guaranteed in
expectation, which makes the endpoint useful for bounding the shared-state claim but insufficient
to identify every multi-agent mechanism. A team learner can still couple private problems through
its training objective.

With partial sharing, some coordinates obey Equation~\eqref{eq:shared-transition-example} and the
rest obey Equation~\eqref{eq:private-transition-example}. This retains interaction while varying a
mechanism that can carry the target-correlation effect. Because the shared coordinates also alter
the credit-assignment problem, a monotonic slope sequence is evidence of modulation, not a clean
mediation estimate. A further control would preserve reward and information coupling while
removing physical contention, or vary credit assignment while holding the transition map fixed.

The worked example also locates authorization. The transition rule determines the operative locus:
joint action aggregation for the shared state, individual action for private states. The rewards
specify modeled objectives. Neither says whether agents authorized the rule, target, action set,
or boundary. A laboratory can test the index--outcome relation without instantiating consent at
all. Any normative application must add the authorization layer rather than infer it from the
ability to coordinate.

Finally, the joint optimum is not necessarily the authorized allocation. The weights $c_i$ may be
engineering coefficients rather than calibrated stakes, and maximizing average reward can accept
large losses for one agent. If the model uses $c_i$ as $s_i$, it must justify the mapping. If a
high-exposure agent withholds authorization, Equation~\eqref{eq:authorization} can fail even when
Equation~\eqref{eq:quadratic-decomposition} has a high optimum. This is exactly the separation
between performance and legitimacy.

\subsection{Interventions by layer}

The formal separation yields different intervention targets. Changing the access-control map
$\operatorname{ctrl}_{\Gamma}(r)$ reallocates operative authority. It can increase or decrease effective voice depending on
who receives control, but the code change itself is not an authorization act. Changing a voting or
delegation rule targets authorization and influence. It may leave objective alignment unchanged.
Changing rewards or preference-learning procedures targets $\alpha$ or $\varepsilon$ under the
measurement model. It may improve performance while leaving control rights untouched. Changing
the transition from shared to private state targets physical contention. It can alter outcomes
without making agents' objectives more similar.

These distinctions matter for causal claims. Suppose a designer introduces a communication
channel and the coordination gap falls. The mechanism could be improved information, implicit
commitment, synchronized exploration, or changed effective control. Measuring only the outcome
and calling the intervention ``more consent'' would not identify any of them. A factorial design
can vary channel capacity while holding rewards, transitions, and authorization rules fixed, then
measure whether the controller actually uses the signal. A separate study can ask whether affected
agents authorized the channel or its controller.

Similarly, reward sharing can change the game. When an algorithm mixes agents' rewards, it may
increase objective correlation by construction, internalize externalities, or transfer costs
between agents. The resulting policy improvement supports the engineered reward mechanism within
that game. It does not show that pre-existing preferences became aligned or that agents consented
to the transfer. The model should report original and transformed rewards and locate which one
defines $\alpha$.

Shared-state factorization provides a third example. Moving from one common resource to private
copies eliminates physical contention only if the copies do not share a capacity constraint or
externality. If a global budget remains, the operative locus may simply move to the budget rule.
The separable MARL control is mathematically clean because rewards and transitions factor at the
optimal-control level; real systems require an audit for hidden coupling. A negative endpoint is
informative only for the interaction actually removed.

Authorization interventions can also change the data-generating process. Giving high-stake agents
veto power may alter which proposals are made, not merely which pass. Consultation may change
preferences through deliberation. Exit rights may change measured stakes by limiting future
exposure. These responses are part of the institution, and a before--after comparison that treats
the proposal set or affected population as fixed can be misleading. The relevant estimand must say
whether it conditions on the offered decisions or includes selection into them.

For design, the practical rule is to attach every intervention to one primary layer and list its
plausible spillovers. Success criteria then follow: access changes are evaluated by realized
influence; authorization changes by verified acts and threshold margins; measurement changes by
construct-validity evidence; coordination changes by separately specified, index-distinct outcomes; and persistence changes
by transition data under the ROM assumptions. No single favorable metric certifies the whole
chain.

\subsection{Outcomes and estimands}

Candidate outcomes include communication overhead, deadlock duration, repeated reallocation,
defection, aggregate reward loss, or distance to an attainable control benchmark. They should be
specified separately from Equation~\eqref{eq:resource-index}, and any shared algebra with its inputs
must be audited. The companion uses
\begin{equation}
\Delta R=R^*-\frac{1}{n}\sum_i\bar R_i,
\qquad R^*=0,
\label{eq:coordination-gap-aoc}
\end{equation}
where $R^*=0$ is the average unconstrained individual ideal under the quadratic reward and
$\bar R_i$ is achieved agent reward. It is not a difference between individual and joint reward.
Joint continuous, reachable-lattice, and dynamic optima enter later as decomposition benchmarks.
Some analyses divide $\Delta R$ by the experimental reward-scale coefficient $\sigma$. The outcome has several versions:
\begin{enumerate}[noitemsep]
\item a late-training window with ongoing updates and exploration;
\item frozen greedy policies on held-out resets;
\item frozen policies evaluated with matched observation noise;
\item frozen policies with the training exploration floor restored; and
\item a residual to an exact finite-horizon dynamic oracle.
\end{enumerate}
These estimands are related but not interchangeable. The training window describes late-training
performance; it mixes policy quality, exploration, updating, and the training reset stream. Frozen
held-out evaluation asks what the learned policy does without further updates. A matched-noise
score adds observation-time error. A dynamic-oracle residual conditions on transition constraints,
reachable states, horizon, and resets. Labeling all of them ``learning friction'' would erase the
nuisances the controls were built to separate.

For the clean frozen policy, the companion decomposes the gap to a continuous joint optimum as
\begin{equation}
R^*-\E R_T^\pi
=\underbrace{R^*-R^{\mathrm{cont}}}_{\Phi}
+\underbrace{R^{\mathrm{cont}}-\E R^{\mathrm{lat}}}_{Q}
+\underbrace{\E(R^{\mathrm{lat}}-R_T^{\mathrm{dyn}})}_{T_c}
+\underbrace{\E(R_T^{\mathrm{dyn}}-R_T^\pi)}_{P}.
\label{eq:oracle-decomposition-aoc}
\end{equation}
Expectations are over the held-out reset bank and any evaluation randomness, conditional on a fixed
target-weight instance; $R^*$ and $R^{\mathrm{cont}}$ are deterministic at that conditioning level.
$\Phi$ is continuous feasibility, $Q$ reachable-lattice cost, $T_c$ the travel transient, and
$P$ a frozen-policy residual to the exact dynamic controller. Even $P$ is not pure learning
friction: for decentralized learners it also contains information and objective mismatches between
individual policies and the joint benchmark. Section~\ref{sec:empirical-status} reports what this
decomposition changes.

\subsection{What this instantiation can test}

The multi-agent application supports a sequence of separable questions:
\begin{enumerate}
\item Does the signed target generator deliver the intended target-vector correlation $\rho$, and
how does that proxy relate to realized reward-function alignment over feasible states?
\item Does the coefficient labeled $\sigma$ represent calibrated exposure rather than merely
rescale the response?
\item Does the information treatment alter learned policies, evaluation-time behavior, or both?
\item Does any coordinate predict $Y$ conditionally, and does the composite improve held-out fit?
\item Does the effect survive frozen held-out evaluation and an attainable benchmark?
\item Is a target-correlation gradient present without shared-state contention, and how does it change as
the shared fraction varies?
\item Do opposition contrasts agree within reward-scale conditions, learners, target supports, and multiplicity
families?
\end{enumerate}
The questions are more durable than the canonical ratio. A failed composite can coexist with an
informative control battery because the battery identifies which treatment, interaction, or
benchmark generated the apparent regularity.

Other applications remain possible, but the paper no longer uses cryptocurrency, political
institutions, or human--AI relationships as parallel demonstrations. They involve different
affected-set rules, utility scales, authorization acts, and outcome processes. Illustrative
mappings and case studies remain in \TSpointer{TS-DOMAIN}; they are hypotheses awaiting construct
validation and measurement-invariance tests, not evidence that one latent friction quantity spans
the domains.

\section{Measurement, falsification, and current empirical status}
\label{sec:measurement}

The framework is testable only if its layers remain operationally distinct. AOC therefore supplies
an identification contract rather than a universal proxy catalogue. The contract states what an
empirical study must fix before fitting an index or interpreting an outcome. Domain-specific
measures can vary; the questions they answer cannot be allowed to drift after results are known.

\subsection{The identification contract}

\begin{enumerate}
\item \textbf{Decision and scale.} Specify the decision, time interval, resource, and scale $S$.
Identify the operative selection procedure, including vetoes, defaults, and aggregation rules.

\item \textbf{Affected set and authorization event.} State how $S_d$ is constructed, which
authorization act is observed, and whether the claim concerns unanimity or a named relaxation
$\theta<1$. Silence, non-exit, policy proximity, and low disruption are not authorization acts by
default.

\item \textbf{Stake calibration.} Declare the allocation contrast $\Delta a$, utility-loss unit
$U_{0,S}$, aggregation rule, and sensitivity to alternative references. Raw monetary exposure,
task criticality, or reward scaling is not automatically commensurable across agents.

\item \textbf{Alignment estimand.} Specify the objective variables and the sampling distribution
over states or actions. Verify the realized sign and range, not merely a generator parameter.
Report undefined constant-objective cases and any clipping or support transformation.

\item \textbf{Information channel.} Identify whose uncertainty about which variable is measured,
what signal is available, and under what distribution. Separate goal direction from predictive
dependence; mutual information and transfer entropy cannot identify signed alignment.

\item \textbf{Index-distinct outcome.} Prespecify $Y$, its normalization, evaluation policy, reset or
sampling distribution, horizon, and benchmark. Audit any algebra shared with the treatment or
index inputs. An input used to compute $F$ cannot simultaneously serve as evidence that $F$
predicts itself.

\item \textbf{Model and comparison.} State the candidate functional, alternatives, interaction
tests, uncertainty estimator, multiplicity family, and held-out criterion. Comparing only the
canonical index with a null model cannot show that its composition is needed.

\item \textbf{Status and provenance.} Mark analyses as confirmatory, prespecified follow-up, or
exploratory. Preserve seeds, configurations, raw outputs, code versions, and the distinction
between local evidence and publicly archived evidence.
\end{enumerate}

Table~\ref{tab:measurement-contract} gives the minimum separation. Extended survey measures,
revealed-preference constructions, bounded divergence transforms, transfer-entropy diagnostics,
proxy catalogues, and causal-design options are retained in \TSpointer{TS-MEASURE}. They can guide
a domain protocol but are not interchangeable validated measures.

\begin{table}[htbp]
\centering
\caption{Minimum measurement contract. ``Failure'' names what a negative result would reject.}
\label{tab:measurement-contract}
\begin{tabular}{@{}p{2.4cm}p{4.4cm}p{6.3cm}@{}}
\toprule
Object & Required declaration & Failure interpretation \\
\midrule
Authorization & affected set, verified act, stake weights, threshold & rejects the claimed
authorization classification, not an outcome model \\
Alignment $\alpha$ & objectives, support $\mu$, signed realization, variance conditions & rejects
the treatment or objective-association measure \\
Stake $\sigma$ & increment, common reference, aggregation, units & blocks cross-agent or
cross-domain comparison; scale effects may be mechanical \\
Information $\varepsilon$ & target variable, signal, conditioning set, estimator & rejects the
channel measure; sign-blind dependence is not alignment \\
Index $F$ & function, normalization, alternatives, fit criterion & rejects the composite if a
simpler or less constrained model predicts better \\
Outcome $Y$ & protocol, horizon, resets, benchmark, noise and exploration & confines the claim to
the estimand actually measured \\
Persistence & duration event, competing risks, survival bridge, ROM conditions & rejects the
descriptive bridge, not the unanimity axiom \\
\bottomrule
\end{tabular}
\end{table}

Measurement error deserves special attention because the ratio is nonlinear. Error in $\alpha$
is amplified near $-1$; denominator noise can create both bias and apparent tail behavior. Error in
$\sigma$ changes scale, and a reward design that is homogeneous of degree one in a parameter
labeled ``stake'' can build the raw main effect into the outcome. Error in $\varepsilon$ can leak information about
policy difficulty or evaluation noise rather than isolate an information channel. Uncertainty
propagation should therefore operate on the primitive measures and repeat the full model
comparison, not attach a standard error only to the final index.

\subsection{Falsification rules}

Four negative results have different meanings.

\paragraph{Construct failure.} If a sampler labeled negative alignment produces nonnegative
realized correlations, the treatment failed. Downstream curves cannot be interpreted as opposition
effects even if they are precisely estimated.

\paragraph{Composition failure.} If $\alpha$, $\sigma$, and $\varepsilon$ have associations with
$Y$ but the ratio predicts worse than an independent-effects model, the composite fails. The
coordinates may remain useful conditionally; their presence does not rescue the specified
denominator or interaction.

\paragraph{Scope failure.} If the alignment association disappears under separable state dynamics
and strengthens as the shared fraction rises, a context-free alignment law fails. The surviving
statement is about contention or shared-coordinate modulation in that environment.

\paragraph{Estimand failure.} If a late-training result changes after policies are frozen,
evaluated on held-out resets, or compared with a dynamic oracle, the original result remains a
late-training finding. It cannot be silently relabeled learned-policy quality.

These rules make nulls informative. A non-detection is not an equivalence result unless an
equivalence margin and test were specified. Opposite-signed estimates pooled across reward-scale
levels cannot support an unconditional null without defining the reward-scale mixture as the estimand. An
exploratory interaction can motivate a follow-up but should not be presented as a preregistered
law. Falsification strengthens the framework only when it changes the claims.

\subsection{A staged analysis plan}

A future confirmatory study can implement the contract in five gates. The first gate is
\emph{treatment delivery}. Before training or outcome analysis, simulate a large draw from the
target and information generators. Report requested and realized signed correlations, boundary
rates, marginal distributions, reward scaling, any separate stake calibration, and channel entropy. Fail the gate if the delivered
treatment is outside a prespecified tolerance. This prevents a nominal parameter from acquiring an
interpretation the environment does not realize.

The second gate is \emph{mechanical benchmarking}. Derive invariances, scale homogeneity, feasible
optima, and attainable dynamic bounds before fitting a learner. A condition whose expected outcome
is algebraically flat is a structural negative control, not an empirical equivalence test. A
condition whose loss scales linearly with an experimental reward coefficient cannot independently
test whether calibrated stakes affect learning. Dynamic oracles should use the same horizon, resets, transitions, action constraints,
and objective as the policy comparison.

The third gate is \emph{estimand registration}. Select a primary outcome, such as clean frozen
held-out policy performance. Register training-window, matched-noise, and exploration-restored
scores as distinct secondary estimands. Specify the reset bank, episode count, randomization unit,
clustering level, learner comparison, and whether normalization changes the scientific question.
This gate prevents a convenient window from being redescribed after results appear.

The fourth gate is \emph{model competition}. Fit the canonical ratio only alongside plausible
simpler and more flexible alternatives. At minimum, compare stakes-only, additive main effects,
pre-stated interactions, and a flexible response surface using held-out likelihood or prediction.
Report absolute fit, not only a coefficient on a generated index. A composite earns support only
if its restrictions improve transport or parsimony without sacrificing predictive adequacy. If an
interaction has a sign implied by the formula, test that sign directly.

The fifth gate is \emph{scope and multiplicity}. Register the primary reward-scale mixture or report
within-scale contrasts. Define the family containing learner, reward-scale, and endpoint comparisons.
Specify equivalence margins if equivalence matters. Separate confirmatory tests from diagnostic
controls and independent replications. A partial-sharing curve should register whether the target
is any modulation, a linear ramp, a threshold, or a monotonic order; those are different
hypotheses.

The companion illustrates why an estimand ledger belongs in the protocol. Its outcome labels can
be organized as follows.

\begin{table}[htbp]
\centering
\caption{Estimand ledger for the MARL battery. Each row removes or restores a specific source of
variation; none is a universal measure of ``coordination.''}
\label{tab:estimand-ledger}
\begin{tabular}{@{}p{2.7cm}p{4.6cm}p{5.7cm}@{}}
\toprule
Estimand & Held fixed or changed & Claim supported \\
\midrule
Late-training gap & updates, training resets, and 1\% random-action exploration remain &
performance during the declared terminal training window \\
Clean frozen gap & parameters fixed; greedy actions; unseen common reset bank; no observation or
action noise & learned-policy behavior on held-out starts \\
Matched-noise frozen gap & clean frozen protocol plus evaluation-time observation noise & combined
policy and observation-time measurement cost \\
Exploration-restored gap & parameters fixed; observation noise matched; 1\% action floor restored &
policy behavior with the execution noise used in training \\
Dynamic-oracle residual $P$ & same starts, horizon, state bounds, lattice, and joint transitions;
best attainable controller subtracted & residual to a matched joint controller, including
decentralization and objective mismatch \\
\bottomrule
\end{tabular}
\end{table}

The ledger explains why superficially similar estimates need not be replications. The archived
late-training analysis and deterministic reviewer rerun use different trained policies and
protocols. The rerun reports both its own late-training outcome and a clean frozen outcome; the
cooperative gradient persists across those two estimands. That comparison does not retroactively
turn the archived CSV into a frozen evaluation of the same policies. Provenance requires keeping
the rows separate. Exact estimates remain in the MARL claim map and its \texttt{TS-FIRST} and
\texttt{TS-COND} records.

Opposition contrasts likewise answer conditional questions. In the archived legacy late-training
design, opposition minus the uncorrelated-target baseline changes sign across reward-scale conditions and none of the
within-scale contrasts survives Holm correction. Pooling those cells would estimate an average over
a specific reward-scale mixture whose opposite-signed components can cancel. In the deterministic
legacy frozen rerun, all four reward-scale-by-learner point estimates are penalty-signed, but only one survives the
declared four-test family. This is neither equivalence nor a replicated universal penalty. The
feasible-centred frozen result reported below answers a new, better-delivered treatment question
while preserving the earlier nulls; exact contrasts and adjusted tests stay in the MARL owner.

The oracle ledger provides a second distinction. $\Phi$ concerns the location of the continuous
joint optimum relative to the feasible box. $Q$ concerns the reset-specific reachable lattice.
$T_c$ concerns finite-horizon travel. $P$ concerns the learned policies relative to the exact
attainable controller. A negative total slope can coexist with a nonnegative $P$ slope when
feasibility carries more than the entire gradient and other components offset it. Reporting only
the sum would invite a learning interpretation that the decomposition rejects.

These gates also improve provenance. Each table row should trace to an immutable configuration,
raw output, analysis script, and exact filter. A manuscript number is not verified by matching a
previous table; it is verified by rerunning the regenerator or recomputing from canonical data.
Build logs, source checksums, code commits, public archives, and local result files are different
evidence surfaces. A data-availability paragraph should name which surface contains which stage of
the record.

\subsection{Measurement invariance and transport}

Even a successful within-domain study would not immediately validate a cross-domain index. The
first transport question is configural: do alignment, stakes, and information describe the same
latent roles in both settings? The second is metric: does a unit change have comparable meaning?
The third is scalar: are levels comparable after normalization? Without these properties, an index
can rank cases within each domain while making cross-domain comparisons meaningless.

Transport can be tested rather than presumed. Estimate the measurement model separately by group,
test whether loadings or calibration functions can be constrained, and examine whether the
index--outcome relation retains sign and shape under leave-one-domain-out evaluation. A failed
invariance test does not make the local measure useless. It blocks the stronger claim that one
number represents the same construct everywhere.

Outcome transport is a separate challenge. Coordination gap, exit, volatility, protest, and policy
reversal have different censoring, time scales, and causal processes. Mapping each to a generic
letter $Y$ is bookkeeping. It does not show that coefficients are comparable or that the same
functional relation should hold. For this reason the current reader paper keeps one worked outcome
family and treats other domains as technical-record illustrations.

\subsection{Validity evidence required for each construct}

The three coordinates need more than internal consistency. Alignment requires \emph{content
validity}: the compared objectives must cover the consequences relevant to affected agents rather
than the subset convenient to log. It also requires \emph{convergent validity}: different credible
objective measures should agree where they purport to measure the same relation. Finally it
requires \emph{discriminant validity}: the alignment measure should not merely recover common task
difficulty, shared scale, or information quality. Signed treatment verification addresses only one
part of this programme.

Stake measurement needs a defensible counterfactual. Exposure is the difference between outcomes
under a declared allocation contrast, not the size or visibility of an agent. A high reward
coefficient can encode designer priority without representing welfare. Monetary exposure can be
observed while nonmarket harm remains omitted. Calibration evidence should include unit invariance,
sensitivity to the contrast and reference, and agreement with external indicators of consequence
where available. If those checks fail, $\sigma$ remains an experimental scaling parameter rather
than a validated stake measure.

Information deficit needs a well-defined source, receiver, and target. Predictive uncertainty about
an agent's next action is not necessarily uncertainty about its utility. A channel can transmit an
accurate preference while the controller ignores it, or transmit little while objectives happen to
align. Interventions should distinguish signal availability, signal use, and evaluation-time
corruption. The MARL battery does this partially by comparing policies trained under observation
noise in clean, matched-noise, and exploration-restored evaluations.

Effective voice $v_i$ needs influence evidence. Self-reported voice may measure perceived inclusion;
formal voting weights measure rules; counterfactual outcome sensitivity measures realized power.
These quantities can disagree and should not be averaged without a measurement model. A controller
can solicit extensive input while retaining complete agenda and veto power. Conversely, a rarely
used veto can carry substantial influence. Authorization records validate a different object again:
whether an act occurred under specified conditions.

Outcome validity is equally important. A reward gap may measure task performance but miss
communication cost. Deadlock time may be zero because a centralized rule always resolves conflicts,
while affected agents lack voice. Exit may be censored by switching costs. Multiple outcomes can be
reported, but selecting the one most favorable to the index after inspection creates another
manufactured finding. A primary outcome and its relation to secondary outcomes should be fixed in
advance.

Common-method bias is a special danger. If rewards define $\alpha$, generate $\sigma$, and enter
$Y$, associations can be mathematical consequences of shared construction. The first-stage raw
reward-scale effect is an example. Remedies include independent exposure measures, outcomes not algebraic
functions of the treatment, negative-control variables, and simulation of the null implied by the
measurement pipeline. A regressor's conceptual name cannot neutralize a shared equation.

Boundary and support audits form another validity class. Correlation specified in latent space may
not survive clipping, truncation, discrete actions, or policy visitation. Report both requested and
realized relations at the latent target level, feasible optimum, and visited state--action support.
The appropriate level depends on the claim. A treatment can be delivered in latent variables yet
attenuated in the decision problem; that attenuation is part of the construct, not random noise to
be ignored.

Finally, validation must be asymmetric with respect to positive and negative findings. A failed
composite should be rejected even if individual variables retain expected signs. A successful fit
should still face alternative functions, held-out tests, and construct checks. Evidence of
non-equivalence requires a detected difference; evidence of equivalence requires a prespecified
margin and appropriate test. Failure to reject is neither. These rules prevent flexible language
from turning every outcome into support.

\subsection{The MARL control battery}
\label{sec:empirical-status}

The companion study \citep{farzulla2026marl} owns the control-battery evidence. AOC needs the
result only to delimit its empirical claims. Most repairs were exploratory responses to diagnosed
artifacts, not a retroactive confirmatory programme. Table~\ref{tab:marl-interface-verdict}
therefore records only the verdicts needed to delimit AOC. Exact estimands, coefficients,
uncertainty, multiplicity families, and model comparisons remain exclusively in the companion's
reader paper, claim/evidence map, and technical supplement; the AOC supplement supplies navigation
rather than a second results report.

\begin{table}[htbp]
\centering
\footnotesize
\setstretch{1.05}
\caption{MARL evidence-status interface. This table carries claim boundaries, not a duplicate
numerical results section. A non-detection is not an equivalence result.}
\label{tab:marl-interface-verdict}
\begin{tabular}{@{}p{2.6cm}p{4.7cm}p{5.6cm}@{}}
\toprule
Artifact & Control and scoped evidence & Consequence for AOC \\
\midrule
Sign-blind treatment and reward scaling & Audit realized rather than nominal correlation; divide
the outcome by its degree-one reward scale before comparing factor rankings. &
Withdraw the U-shape and quadratic motivation. Neither raw reward-scale dominance nor the normalized
ranking validates a structural importance order. \\

Contention and support & Contrast feasible-centred separation with shared state, and vary overlap
on paired legacy-Gaussian-at-zero support. IQL is structurally invariant at the feasible separable
endpoint; the VDN slope is not detected. & A cooperative target-correlation gradient appears in the
tested shared-state family. Four overlap levels show monotonic point-estimate modulation, not an
exact dose law, VDN equivalence at separation, or a unique causal mechanism. \\

Evaluation and benchmark & Freeze policies, use held-out resets and feasible-centred targets, and
compare the total gap with an exact finite-horizon oracle under matching dynamics. &
The static gap was not pure learning shortfall. Both feasibility and residual-policy channels can
matter, and their attribution is target-support dependent; the effect is not ``entirely
feasibility.'' \\

Composite and proxy channels & Compare the ratio with separately estimated terms, then cross signed
target correlation with observation noise under feasible-centred frozen evaluation. Under this
proxy instantiation, the independent model fits better and the interaction runs opposite to the
ansatz. & The composite fails as a predictor in the tested proxy/outcome design, and the proxy
interaction has the opposite qualitative direction. A positive average noise penalty does not
imply a penalty in every reward-scale--target-correlation cell. \\
\bottomrule
\end{tabular}
\end{table}

For AOC, the battery separates four questions that the long edition had allowed to blur. A nominal
treatment label is not a measure of delivered objective alignment; reward units are not evidence
of structural importance; a late-training gap is not a policy-quality estimand; and correlation
among candidate coordinates does not validate their proposed one-dimensional composition. The
companion manipulates cross-agent \emph{target-vector} correlation, not the general reward-function
correlation in Equation~\eqref{eq:marl-alignment-reader}. Connecting the two requires a declared
state distribution, reward weights, clipping rule, and visitation bridge. Its observation-noise
treatment is likewise one intervention on one information channel, not a domain-general measure of
epistemic access. Finally, the frozen coordination gap remains an outcome under a particular reset
bank, horizon, action lattice, learner, and reward family. It is neither consent nor legitimacy.

Support and inferential scope remain load-bearing. The partial-sharing series uses paired legacy
Gaussian-at-zero targets, whereas the standalone separable control is feasible-centred; their
endpoint estimates need not coincide. The endpoint smoke gate matches reported summaries, not
internal execution traces or full-run equivalence. Feasible-centred opposition penalties appear at
both tested reward-scale levels and learners, but reward-scale moderation is not detected; the result is not evidence
for a universal adversarial singularity. Separable VDN remains a non-detection, not an equivalence
finding. All exact estimates, adjusted tests, legacy contrasts, and null families stay in the MARL
owner rather than being restated here.

The current evidence is traceable through the companion Zenodo v3.0.0 release inventory
(\href{https://doi.org/10.5281/zenodo.22004215}{10.5281/zenodo.22004215}). Repository and archive
states predating the synchronized August repair do not contain the complete package and are
insufficient for reproduction. Exact filters and source files are recorded only in the MARL
claim/evidence map and companion record.

\subsection{Verdict for AOC}

The empirical verdict has three levels.
\begin{enumerate}
\item \textbf{Failed in exploratory comparisons under the tested proxy/outcome instantiation:} the
canonical composite as a single predictor; the qualitative target-correlation--observation-noise
interaction encoded by that instantiation; the sign-blind U-shape; the quadratic refinement as an
empirical claim; reward-scale dominance as evidence about learning; and a universal adversarial
singularity.

\item \textbf{Observed in exploratory controls within scope:} a cooperative target-correlation gradient in shared-state
resource contention; monotonic point-estimate strengthening across four paired legacy-support shared-coordinate
fractions; a feasible-centred opposition penalty at both tested reward-scale levels and learners; a positive
average policy cost of observation-noise training; and two support-dependent components of the
feasible-centred gradient, continuous achievability and frozen-policy residual.

\item \textbf{Unresolved or separate:} equivalence at the separable VDN endpoint; a unique causal
mechanism for partial-sharing modulation; generalization beyond the tested target supports,
learners, horizons, and reward family; empirical adequacy of $L$; and the survival bridge or ROM
stationary ranking.
\end{enumerate}

This verdict changes what AOC may claim. The architecture survives because it made the failed
composition and the index-distinct outcome link visible. The formula does not survive as a
validated law in this proxy instantiation. Alignment remains a candidate directional variable only
with contention and support scope conditions. Information loss is a separate channel whose
interaction should be estimated, not inherited from an observation-noise proxy. Stakes require
calibration and algebraic auditing. No observed low gap
licenses an inference of consent.

The companion record contains the August controls and code paths underlying these numbers. Zenodo
v3.0.0 publishes the synchronized data, analysis, inventory, and validation record; repository and
archive states predating that repair are insufficient for reproduction. The empirical claims
should therefore be checked against the companion release inventory and checksums.
\TSpointer{TS-STATUS} preserves the status chronology; the companion paper remains the
authoritative location for estimands, tables, code mapping, and multiplicity details.

\section{Objections, scope conditions, and adjacent theories}
\label{sec:scope}

The layer separation resolves some objections by refusing the inference they target. Others expose
genuine commitments that the framework cannot settle internally. This section states the strongest
versions because the architecture is useful only if it remains clear where authorization ends and
additional theory begins.

\subsection{Low observed friction can coexist with coercion}

An authoritarian or technically constrained system can display little protest, exit, or measured
coordination loss despite deep opposition. Repression can raise the cost of expression; surveillance
can prevent organization; dependency can block exit; and a measurement instrument can miss latent
disagreement. The framework must not answer this case by defining an unobserved stock large enough
to save every prediction. That would make the theory unfalsifiable.

The defensible response is measurement separation. Let $Y^{\mathrm{obs}}$ be an observed outcome
and $Z$ a vector of expression or attenuation mechanisms. A study may posit
\begin{equation}
Y^{\mathrm{obs}}=h(Y^{\mathrm{latent}},Z)+u,
\label{eq:attenuation}
\end{equation}
but $h$ and $Z$ require independent indicators or an identification design. Neither high $F$ nor
low $L$ proves that latent resistance exists. Conversely, low $Y^{\mathrm{obs}}$ does not prove
authorization. The long technical record \TSpointer{TS-OBJECTIONS} develops suppression and
unexpressed-resistance examples; they are proposed outcome models, not devices for immunizing the
index.

The same point applies in computational systems. A low coordination gap can be produced by a
centralized optimizer that ignores agent voice, or by a reward structure that makes disagreement
costless. A high gap can arise among fully authorized agents facing a difficult search problem.
Outcome quality, authorization, and legitimacy should be reported as separate columns rather than
collapsed into a single favorable score.

\subsection{Consent quality exceeds a binary act}

The authorization predicate records an act under a stake-weighted rule. It does not by itself show
that the act was informed, competent, uncoerced, revocable, or selected from an acceptable option
set. These are not minor measurement details; they are constitutive questions for consent. An
institution could satisfy a recorded vote while manipulating the agenda or withholding material
information. A model could elicit a preference after shaping the available responses.

A complete normative account would add conditions on information, capacity, coercion, and options.
AOC leaves them as explicit scope conditions rather than pretending that $\varepsilon$ solves the
problem. The information coordinate measures a declared controller-side uncertainty channel. It
does not certify that affected agents were informed, that communication was truthful, or that the
decision was free. Likewise, $\alpha$ measures objective association, not voluntariness.

This limitation also constrains delegation. An agent may authorize a procedure rather than each
decision. The relevant decision locus then includes the delegation rule, and the authorization
event concerns that rule within its declared scope. Consent to one domain cannot be presumed to
transfer to another; persistence of the delegation does not establish continuing authorization;
and revocation costs belong in the institutional description.

\subsection{Unanimity may be infeasible}

The Axiom's unanimity boundary invites the familiar objection that collective action would be
paralyzed by holdouts, strategic vetoes, or urgent decisions. The objection is serious. The
framework's answer is not to redefine majority rule as consent. It is to report the trade-off
honestly. A procedure may use $\theta<1$ because unanimity is costly or impossible, while remaining
a relaxation that binds some affected parties without their authorization under the axiom.

This distinction permits institutional comparison. One can measure delay, bargaining cost,
distributional harm, and effective voice under different thresholds. One can add rights or
emergency constraints. One can ask whether compensation or exit changes affected stakes. None of
those analyses requires semantic inflation. The axiom supplies a normative reference point; an
institution supplies an implementable rule; empirical work measures consequences.

Strategic stake inflation creates a related problem. If parties can manipulate $s_i$, a
stakes-weighted threshold can reward exaggeration. Calibration must therefore rely on auditable
exposure measures or mechanism-specific elicitation, and may need caps, verification, or robust
aggregation. AOC does not solve strategic elicitation. Mechanism design is the appropriate
adjacent toolkit for incentive-compatible reporting
\citep{hurwicz1960optimality,myerson1981optimal,maskin1999nash}.

\subsection{The framework does not bridge is and ought}

An arrangement can be evolutionarily stable and morally objectionable. The survival bridge in
Equation~\eqref{eq:survival-bridge} is descriptive once stipulated; the authorization principle is
normative. Connecting them requires a bridge premise, not a theorem. The framework adopts no claim
that differential persistence makes a configuration right. It also rejects the inverse: morally
authorized arrangements need not persist.

The limited normative proposal is this: affected agents' authorization, weighted by calibrated
stakes, matters for legitimacy. The empirical programme asks whether voice, objective alignment,
information, and exposure predict separately specified, index-distinct outcomes after shared
algebra has been audited. Evidence for those associations
can inform institutional design, but cannot by itself establish the axiom. Evidence against the
canonical index can revise the descriptive model without revising the normative commitment.

This division differentiates the proposal from evolutionary ethics. Evolutionary game theory
explains strategy frequencies and convention selection
\citep{maynard1973logic,skyrms1996evolution,young1993evolution}. AOC uses an evolutionary interface
only after specifying an effective-voice variable, its declared legitimacy bridge, and a survival
map. The addition makes the normative
commitment visible; it does not turn the model into a proof of that commitment.

\subsection{Alignment is not agreement, welfare, or consent}

Correlation between objectives is a coarse relation. It ignores levels, curvature, tail harms,
and rights. Perfect positive correlation can coexist with very different utility magnitudes;
zero correlation can hide local agreement on the states a policy actually visits; negative
correlation over a broad support can coexist with agreement in the feasible region. Aggregation can
mask a high-stake opposed minority.

These limits explain why the companion result cannot be restated as ``cooperation is legitimate.''
Its exploratory controls estimate a signed target-coordinate relation for a specified reward-gap
outcome under a particular shared-state MDP. They contain no authorization act and no empirical
measurement of $L$. The surviving gradient is a scope-conditioned construct-validity result about
contention and target feasibility, not a moral result. Any human
or institutional application would need separate evidence about agency, affectedness, and
authorization.

For similar reasons, information can have negative social value in coordination games. More
precise public signals can worsen welfare when agents overweight common information
\citep{morris2002social}, and transparency about actions can distort incentives
\citep{prat2005wrong}. The positive $\partial F/\partial\varepsilon$ is a property of the ansatz,
not a general theorem that more information always improves coordination. In the companion's
reduced feasible-centred frozen-policy crossing, training with observation noise raises the mean
reward gap while its exploratory interaction with target correlation runs in the opposite direction
from the ansatz. That proxy result does not validate a general entropy claim.

\subsection{Scale mixing can reverse conclusions}

The affected set and decision locus are scale-dependent. A team may authorize an internal policy
that imposes costs on outsiders. An institution may aggregate local majorities into a global
outcome rejected by most high-stake agents. A model trained on individual rewards may be evaluated
against a team objective. Without a scale transformation, the same symbol $L$ or $F$ can refer to
different objects.

Exact lumpability is one sufficient condition for a closed coarse dynamic, but it is not a generic
license to aggregate. When fine types within a block have different effective flows, projected
dynamics can depend on hidden composition and history. Empirical work should therefore prefer a
declared scale and test robustness to plausible affected-set and aggregation choices. A cross-level
comparison needs measurement invariance, not just normalized variables.

Cross-domain generalization is a stronger version of the same problem. Monetary exposure,
computational task criticality, and political affectedness do not become commensurable because each
is mapped to $[0,1]$. A domain-specific normalization can support within-domain ranking while
destroying between-domain meaning. The current paper makes no claim that one calibrated index can
rank a resource-allocation experiment, a public institution, and a market event on a common scale.

\subsection{Causal identification is additional work}

The identification contract distinguishes measurement from causal effect. Even a well-measured
association between $\alpha$ and $Y$ can reflect common task difficulty, selection into governance
forms, or feedback from outcomes to objectives. Stakes may determine both control rights and
exposure. Information interventions may simultaneously change incentives. A regression of $Y$ on
$F$ cannot untangle these channels merely because the formula labels them.

Randomized interventions are preferable when possible. In simulations, signed targets, support,
sharing, observation noise, and evaluation protocols can be crossed or paired. In institutions,
instrumental variables, discontinuities, difference-in-differences, or synthetic controls may be
appropriate only when their assumptions are defended. The technical record
\TSpointer{TS-MEASURE} catalogues such designs, but a catalogue is not an identification result.
Transfer entropy can diagnose incremental lagged prediction; it cannot supply causal sufficiency or
the sign of alignment.

Prespecification matters especially after an artifact is found. The original factorial and its
planned tests can remain confirmatory relative to their protocol, while controls designed after
diagnosis are exploratory. Stronger evidence would come from an independent replication of the
repaired signed, feasible, frozen design with a multiplicity plan fixed before outcomes. Repository
timestamps and dated protocols help reconstruct chronology but do not retrospectively
preregister a test.

\subsection{Adjacent theories divide the labor}

\paragraph{Mechanism design.} Mechanism design asks which rules induce truthful revelation or
efficient allocation. AOC asks which agents have standing, whether the operative rule was
authorized, and whether a separately specified outcome follows the proposed index after common
construction has been audited. Incentive
compatibility can improve the measurement of stakes or preferences but does not settle the axiom.

\paragraph{Social choice and contractualism.} Social choice clarifies aggregation constraints;
contractualism supplies richer accounts of reasonable rejection
\citep{scanlon1998we}. AOC contributes an actual-authorization boundary and a measurement
architecture. It does not solve Arrow-type trade-offs or define reasonableness.

\paragraph{Principal--agent theory.} Delegation, monitoring, and residual loss motivate why
alignment and information might affect coordination cost. The long-form agency model is retained
in \TSpointer{TS-FORM}. It motivates directional ingredients but does not uniquely derive the
multiplicative ratio; decision-relevant utility loss is not interchangeable with information
measured in bits.

\paragraph{MARL and cooperative AI.} Reward sharing, value decomposition, and learned cooperation
study how policies coordinate under specified games \citep{sunehag2018value,gemp2022d3c}. They
provide experimental tools and alternative mechanisms. AOC's distinctive role is to keep reward
alignment, operative control, authorization, and observed performance from being treated as the
same construct.

\subsection{Formal verification limits}

The Lean 4 artefacts verify algebraic statements about the canonical formula: non-negativity on
the declared domain, derivative signs, boundary values, and related identities. They do not verify
that $\alpha$, $\sigma$, or $\varepsilon$ are valid constructs; that the normalization is
appropriate; that $F$ predicts an outcome; that a decision was authorized; or that ROM's bridge is
empirically true. The code also does not formalize the full philosophical argument or the entire
dynamic system. The theorem inventory and listings are retained in \TSpointer{TS-LEAN}.

Formal verification is therefore internal assurance, not external validation. A proof assistant
can establish that the selected ansatz has the claimed monotonicity while the companion's
exploratory proxy battery does not support its composite and cross-effect as descriptions of the
specified reward-gap outcome. There is no contradiction: the proof concerns the equation; the
experiment concerns whether the equation describes that measured outcome.

\subsection{Design implications after the composite failure}

Rejecting the ratio changes institutional advice. It would be premature to optimize
Equation~\eqref{eq:index-allocation} or rank systems by $F$. A design process should instead report a
dashboard of primitive constructs and outcomes: who controls each resource, which affected agents
authorized the rule, how stakes were calibrated, how objective relations vary over feasible
states, what information channels are available, and how policies perform under matched
benchmarks. A dashboard is less elegant than a scalar, but it preserves the channels the evidence
shows can interact differently.

The partial-sharing result suggests a concrete engineering question: where do agents physically
contend over the same state, and can some coordinates be separated without destroying necessary
coordination? Factorization may reduce one source of alignment-sensitive conflict. Yet separation
can duplicate resources, ignore externalities, or remove beneficial pooling. The control battery
does not imply that private state is normatively superior or globally efficient. In the exploratory
target-correlation proxy series, the point-estimate gradient changes with the shared-coordinate
fraction. That modulation motivates shared-state transformation as a mechanism to test; it does
not identify a unique causal mechanism.

Information interventions should likewise be evaluated as channels, not assumed improvements.
Better preference elicitation may reduce controller uncertainty, while additional public signals
can create correlated errors or strategic disclosure. Interpretability can reduce an observer's
uncertainty about a controller without reducing the controller's uncertainty about affected
agents. These are two directions of information flow. An extended model should name both rather
than put ``transparency'' into one $\varepsilon$.

Authorization design remains independent of those engineering choices. Separating a resource can
change the affected set and operative locus; it does not retroactively authorize the original
rule. Improving predictive alignment can reduce a coordination gap; it does not substitute for an
authorization act. Adding a consultation channel can increase formal voice while leaving agenda
control untouched. Each intervention should state which layer it targets and which outcome would
show success.

The framework also recommends a publication discipline. Report failed treatments and rejected
forms in the main text, not only in an appendix. Preserve old estimands with accurate names rather
than replacing them with stronger ones. When a control changes the answer, present the discrepancy
as evidence about construct sensitivity. Keep public archives synchronized with the claims readers
are asked to verify. These practices are methodological consequences of layer separation: no
surface should quietly stand in for another.

\subsection{Scope statement}

The current framework is best read as an authorization architecture plus a disciplined empirical
interface for bounded multi-agent decision domains. It is applicable when operative control and an
affected set can be identified, stakes can be calibrated within a declared scale, objective and
information measures have valid support, and outcomes can be separately observed with any shared
algebra audited. ROM adds
persistence claims only under its stated finite, static, density-independent conditions and a
specified transformation kernel.

Outside those conditions, the notation may still organize questions but does not deliver a
prediction. In particular, the paper establishes no universal convergence toward consent, no
cross-domain scalar of legitimacy, no inference of authorization from low observed friction, no
general adversarial singularity, and no validated composite index. Those negative boundaries are
part of the result.

\section{Conclusion}
\label{sec:conclusion}

The Axiom of Consent is a stake-weighted unanimity principle. It is not a claim that every system
uses unanimity, and it is not weakened into majority rule by changing a parameter. The threshold
model makes the boundary explicit: $\theta=1$ formalizes the axiom; $\theta<1$ represents a named
institutional relaxation. Before either can be assessed, the operative decision locus and affected
set must be identified. Control is structural. Authorization is additional.

Around that normative rule, the paper has separated four empirical objects. Effective voice $L$
is a measurement of stakes-weighted influence. Alignment, calibrated stake, and information
deficit are candidate coordinates. The canonical $F=\sigma(1+\varepsilon)/(1+\alpha)$ is a
phenomenological ansatz. An observed outcome $Y$ is specified separately and audited for shared
algebra with the index inputs. No link among these
objects is definitional. A low outcome cannot establish consent; a favorable objective
correlation cannot establish authorization; and a theorem about the selected formula cannot
establish its empirical adequacy.

The ROM interface adds an equally conditional claim. If configurations evolve through a finite,
static, density-independent continuous-time replicator--mutator process; if persistence is
stipulated as a function of effective voice interpreted through a declared legitimacy bridge and
normalized model friction; and if the fixed positive operator is irreducible, then ROM's restricted
stationary-composition and convergence result follows. Aperiodicity is an additional condition for
normalized discrete-time power convergence, not for that continuous-time result. This is a
persistence model, not a derivation of legitimacy from survival. Scale
transitions, lumpability, ownership, and
identification remain ROM's formal territory rather than parallel contributions of AOC.

The multi-agent application shows why these boundaries matter. In the companion's exploratory
target-correlation/observation-noise proxy instantiation and reward-gap outcome, the composite
fails the model comparison and the proxy interaction has the opposite direction from the ansatz.
The experiment observes neither an authorization act nor an empirical value of $L$, so these
results cannot validate the authorization layer or a legitimacy interpretation.
The initial U-shape was sign-blind, the raw reward-scale effect was partly built into reward algebra,
and the learned-policy interpretation required frozen held-out evaluation and an attainable dynamic
oracle. What survives is narrower: cooperative target correlation lowers the tested gap under shared-state contention;
in a paired legacy-Gaussian-support series, the point-estimate gradient strengthens across four
shared-coordinate fractions; and target
support changes both opposition and residual-policy conclusions. At the fully separable endpoint,
IQL is structurally invariant and VDN yields a non-detection, not equivalence. Feasible-centred
opposition penalties survive within both sampled reward-scale levels and learners, while legacy results remain
mixed and protocol-bound.

The surviving contribution is therefore not a universal friction law. It is a way to keep
authorization, control, modeling, and outcomes from borrowing evidential force from one another.
That architecture is stronger after the composite's failure: it says exactly which claim was
rejected and which questions remain open. Future work should begin from the repaired measurement
contract, preregister an independent signed and feasible replication, compare functional forms on
held-out outcomes, and test effective voice and persistence as separate constructs. It should not
restore the ratio by relabeling individual directional effects as validation.

The same discipline applies beyond MARL. A new application must earn its affected set, stake
calibration, authorization measure, and outcome link within that domain before any cross-domain
comparison is attempted. Formal similarity is a proposal for testing, not transport evidence.
Likewise, a persistence analysis must report the type partition, transition opportunities, and
mutation kernel rather than cite the restricted ROM theorem by analogy. These requirements make
the programme slower but keep its normative and empirical claims auditable.

Consent remains normatively important even when a convenient empirical index fails. Conversely,
a useful predictor would not make consent reducible to low resistance. The framework's most durable
claim is the firewall between those propositions.

\section*{Acknowledgements}
This paper benefited from collaboration with Claude (Anthropic) on formal-verification support,
literature synthesis, and iterative manuscript development, and from adversarial cross-review by
ChatGPT (OpenAI) and OpenAI Codex on mathematical consistency, source-level revision, and build
validation. The author reviewed the resulting text and takes responsibility for all claims,
judgments, and errors.

\section*{Declarations}
\paragraph{Conflict of Interest.} The author declares no competing interests.

\paragraph{Funding.} This research received no external funding.

\paragraph{Data and code availability.} The Lean 4 formalizations are available at
\url{https://github.com/dissensus-ai/lean-formalizations}. The companion MARL study's superseded
public predecessor remains Research Square v1
(\href{https://doi.org/10.21203/rs.3.rs-10375465/v1}{doi:10.21203/rs.3.rs-10375465/v1}).
The current code is at \url{https://github.com/dissensus-ai/friction-marl}; the versioned data,
analysis, inventory and validation record supporting the present evidence summary is Zenodo v3.0.0,
\href{https://doi.org/10.5281/zenodo.22004215}{10.5281/zenodo.22004215}, under concept DOI
\href{https://doi.org/10.5281/zenodo.20636467}{10.5281/zenodo.20636467}. That record contains seven
files. Archive and repository states predating the synchronized August 2026 repair do not contain
the complete record summarized here. The release inventory and checksums therefore form part of
the evidence trail. The
checksum-identified long technical record remains preserved separately from this reader-facing
version.

\paragraph{AI assistance.} Claude (Anthropic), ChatGPT (OpenAI), and OpenAI Codex were used for
the tasks described in the acknowledgements. The author reviewed and approved the manuscript and
takes sole responsibility for its contents.

\bibliography{references}
\end{document}